\documentclass[aps,pre,amsmath,amsfonts,floatfix,superscriptaddress,nofootinbib]{revtex4}

\usepackage{mathrsfs}
\usepackage{t1enc,latexsym,amsfonts, amssymb, amsmath,empheq,amsthm}
\usepackage{graphicx,color}
\usepackage{bbm}
\usepackage{multirow}
\usepackage{mathrsfs}
\usepackage{commath}
\usepackage{hyperref}

\usepackage[french,english]{babel}
\usepackage[utf8x]{inputenc}
\usepackage[T1]{fontenc}
\usepackage{times}
\usepackage{url}

\font\msytw=msbm9 scaled\magstep1

\let\a=\alpha \let\b=\beta \let\g=\gamma \let\d=\delta
 \let\z=\zeta \let\h=\eta 
\let\l=\lambda \let\m=\mu \let\n=\nu  
\let\s=\sigma \let\t=\tau  
   \let\G=\Gamma

  \let\th=\theta \let\io=\infty

\def\MM{{\cal M}} 
 \def\HH{{\cal H}}
 
\def\LL{{\cal L}}  \def\OO{{\cal O}}
\def\DD{{\cal D}} 
\def\KK{{\cal K}}  
\def\ZZ{{\cal Z}}

\def\bth{\bar\th}

\def\de{\mathrm d}

\def\to{\rightarrow} \def\la{\left\langle} \def\ra{\right\rangle}

\def\SSS{\hbox{\msytw S}}

\newcommand{\beq}{\begin{equation}} \newcommand{\eeq}{\end{equation}}
 \newcommand{\wt}{\widetilde}

\usepackage[usenames,dvipsnames]{xcolor}

\usepackage{color}
\definecolor{darkgreen}{rgb}{0,0.6,0}
\definecolor{darkblue}{rgb}{0,0,0.6}
\definecolor{darkred}{rgb}{0.6,0,0}
\definecolor{darkpurple}{rgb}{0.5,0,0.5}

\hypersetup{
bookmarksopen=true,
pdftitle="",
pdfauthor="Elisabeth Agoritsas", 
pdftoolbar=false, 
pdfstartview={FitH},		
pdfmenubar=true,			
pdfhighlight=/O,			
colorlinks=true,			
urlcolor=darkblue,
citecolor=darkblue,		
linkcolor=darkpurple	
}

\newcommand{\argc}[1]{\left[#1\right]}
\newcommand{\arga}[1]{\left\lbrace #1\right\rbrace }
\newcommand{\argp}[1]{\left(#1\right)}

\newcommand{\moy}[1]{\left\langle  #1 \right\rangle }

\renewcommand{\vec}[1]{{\bf #1}}
\newcommand{\ththbar}[1]{\theta_{#1} \bar{\theta}_{#1}}

\definecolor{shadecolor}{RGB}{220,220,220}

\begin{document}

\title{
Out-of-equilibrium dynamical mean-field equations
for the perceptron model
}


\author{Elisabeth Agoritsas\footnote{Email: elisabeth.agoritsas@lpt.ens.fr}}
\affiliation{Laboratoire de physique théorique, Département de physique de l’ENS, École normale supérieure, PSL Research University, Sorbonne Universités, CNRS, 75005 Paris, France}


\author{Giulio Biroli}
\affiliation{Institut de physique th\'eorique, Universit\'e Paris Saclay, CNRS, CEA, F-91191 Gif-sur-Yvette, France}

\author{Pierfrancesco Urbani}
\affiliation{Institut de physique th\'eorique, Universit\'e Paris Saclay, CNRS, CEA, F-91191 Gif-sur-Yvette, France}

\author{Francesco Zamponi}
\affiliation{Laboratoire de physique théorique, Département de physique de l’ENS, École normale supérieure, PSL Research University, Sorbonne Universités, CNRS, 75005 Paris, France}


\begin{abstract}

Perceptrons are the building blocks of many theoretical approaches to a wide range of complex systems, ranging from neural networks and deep learning machines, to constraint satisfaction problems, glasses and ecosystems.
Despite their applicability and importance, a detailed study of their Langevin dynamics has never been performed yet.
Here we derive the mean-field dynamical equations that describe the continuous random perceptron in the thermodynamic limit, in a very general setting with arbitrary noise and friction kernels, not necessarily related by equilibrium relations.
We derive the equations in two ways: via a \emph{dynamical cavity method}, and via a path-integral approach in its supersymmetric formulation.
The end point of both approaches is the reduction of the dynamics of the system to an effective stochastic process for a representative dynamical variable.
Because the perceptron is formally very close to a system of interacting particles in a high dimensional space, the methods we develop here can be transferred to the study of liquid and glasses in high dimensions.
Potentially interesting applications are thus the study of the glass transition in active matter,
the study of the dynamics around the jamming transition, and the calculation of rheological properties in driven systems.

\end{abstract}


\maketitle

\begin{center}
\end{center}


\begin{center}
\rule{200pt}{0.5pt}
\end{center}

\tableofcontents
%

\clearpage

\section{Introduction}
\label{sec-introduction}

A large class of statistical systems, \textit{i.e.}~systems made up of a large number of degrees of freedom, display a complex behaviour because their dynamical evolution takes place in a very rough and high dimensional energy landscape full of minima, maxima and saddles.
The prototype of such systems are glasses.
At the mean-field level, a dynamical theory of glass dynamics has been developed, both in equilibrium and out-of-equilibrium~\cite{Bouchaud,Cu02,CC05}. 
This theory has been extremely successful, having been applied to equilibrium dynamics~\cite{SZ81,KW87,KT87,KT88}, out-of-equilibrium dynamics
in the aging~\cite{CK93,CK94} and driven~\cite{BBK00} regimes, and in presence of particle self-propulsion as a model for active matter~\cite{BK13}.
A general picture of how the rough energy landscape influences the dynamics (called the ``Random First Order Transition'', or RFOT, picture) 
has emerged from these studies~\cite{KW87,KT87,KT88,KTW89}: this picture is exact in the mean-field limit,
and it has been the source of a lot of inspiration for the study of systems outside mean-field~\cite{Ca09,BB11,WL12}.

Despite these successes, several problems remain open. The theory has been developed using mainly a class of toy models,
the so-called ``spherical $p$-spin glass models'', whose dynamical equations also correspond to the so-called schematic limit of the Mode-Coupling Theory equations~\cite{RC05,CC05}.
While these models share most of the basic phenomelonogy of real glasses~\cite{KT87,KT88}, it remains highly desirable
to develop a theory specific to particle systems.
This has been achieved recently by considering the infinite-dimensional limit~\cite{MKZ16a,MKZ16b,CKPUZ16,Sz17},
but only in the special case of equilibrium dynamics.
The extension to the out-of-equilibrium regime is yet to be done but it would have potentially very interesting applications, for example:
{\it (i)} the study of the glass transition in active matter, where the behaviour of the glass transition line is seen to depend on the details of the interaction potential~\cite{Be14,FSB16};
{\it (ii)} the dynamical scaling in the vicinity of the jamming transition~\cite{OT07,Ol10,DDLW15,KCIB15}; {\it (iii)} the dynamical behaviour in rheological experiments,
where several interesting phenomena such as plasticity, yielding, and non-Newtonian flow curves appear~\cite{IBS13}.
These phenomena cannot be fully captured by the simplest $p$-spin glass models. 
Extensions of the Mode-Coupling Theory --~which, as said above, in the so-called `schematic' limit correspond to the $p$-spin dynamics~-- to out-of-equilibrium situations have been obtained.
These extensions obtained partial successes in the flow~\cite{FC02,Br09}, active~\cite{Sz16}, and aging~\cite{La00} regimes, but failed in other cases~\cite{IB13}, calling for alternative approaches.

While our ultimate goal is the study of particle systems, here we take a first step by studying the dynamics of a prototype model, called the ``spherical random perceptron'',
which is somehow intermediate between spherical $p$-spin glasses and particle systems. This model shares most of the basic phenomenology of $p$-spin glasses (in particular the RFOT phenomenology), but it also displays additional interesting features that characterise particle systems, such as a Gardner and a jamming transition~\cite{FP16, FPUZ15, FPSUZ17}.
Perceptron models, 
introduced by McCulloch and Pitts as simple models of a neuron~\cite{McP43},
and later proposed by Rosenblatt~\cite{Ro58} as the simplest unit of a learning machine~\cite{Ni08,gyorgyi},
appear \textit{mutatis mutandis} as the building blocks of many theories in a broad range of 
different fields of science.
The perceptron problem can be seen as the simplest classification task: given a set of inputs or patterns and a set of associated outputs, 
one wants to find the \emph{synaptic weights} such that the input patterns are correctly classified as the prescribed outputs.
More complicated constructions can be developed as soon as different perceptrons are glued together in a feed-forward network which is the simplest versions of modern deep neural networks \cite{LBH15}:
such multilayer generalisations are different ways to take into account non linearities in the classification tasks.

This classification task can also be seen as a constraint satisfaction problem.
Indeed each input pattern should match the corresponding output signal.
Therefore, one should determine a configuration of the synaptic weights which satisfies simultaneously all the input-output constraints.
In this setting, the {\it capacity} of the perceptron is defined as the total volume, in the phase space of the synaptic weights, that are compatible with the constraints.
The {\it random} perceptron is the case in which the inputs and the associated outputs are taken as uncorrelated random variables; in this case the problem is not a machine learning problem, because even if all the input-output relation are correctly classified, it is impossible to generalise to new input-output patterns.
Still, the problem is well-defined as a constraint satisfaction problem, and the computation of the associated capacity has been performed in the pioneering works of Derrida and Gardner~\cite{Ga88, GD88}.

Very recently, in the constraint satisfaction setting, the perceptron has received a renewed interest.
Indeed, using continuous spherical variables, one can build a continuous non-convex constraint satisfaction problem,
for which the limit of zero capacity corresponds to the satisfiability/unsatisfiability (SAT/UNSAT) threshold where the volume of solutions for the compatible synaptic weights shrinks continuously to zero.
This SAT/UNSAT transition is analogous to the jamming transition of hard-sphere glasses in very high dimensions~\cite{FP16, FPUZ15, FPSUZ17},
being characterised by the very same critical exponents of amorphous packings of hard spheres~\cite{CKPUZ14NatComm}.
Therefore, the perceptron is the simplest toy model that contains the universal mean-field physics of glasses and jamming.
Finally, variants of this model have appeared recently to study models of ecosystems: in this case the limit of zero capacity represents the point where the ecosystems reach a stationary equilibrium state~\cite{TM17}.

Hence, the \textit{perceptron cornucopia}, that started in the fields of machine learning and neural networks, now extends to many other research fields, from
glasses to ecosystems.
Despite this wide range of applications, and despite early attemps at a full characterisation of the dynamics~\cite{Ho92},
a complete derivation of the dynamical equations of the perceptron in a broad out-of-equilibrium setting has not been reported.
In this work we study the Langevin dynamics of the spherical random perceptron, 
namely a simple stochastic gradient descent dynamics for the perceptron degrees of freedom, 
with an additional noise term (see Sec.~\ref{sec-def-model} for details).
The dynamical mean-field equations (DMFE) that we derive are very general:
they include general non-local friction terms, general types of colored stochastic noise, and external driving forces, thus allowing for the study of all possible equilibrium and out-of-equilibrium dynamical regimes.
As mentioned above, this study paves the way to derive and solve the out-of-equilibrium DMFE of particles in high dimensions, which, however, we leave for future work.

For pedagogical reasons, and for the sake of future generalisations, we present two distinct derivations of the DMFE: in the first one we use a \emph{dynamical cavity approach}~\cite{MPV87} that is extremely powerful and intuitive (Sec.~\ref{sec-cavity-method}).
In the second one we make use of the \emph{Martin-Siggia-Rose-Janssen-De Dominicis formalism}~\cite{MSR73,Ja76,DD78} 
for path integrals in its supersymmetric version~\cite{Ku92,Cu02} (Sec.~\ref{sec-SUSY}) and we show that both approaches lead to the same results.
Through both approaches, the mean-field perceptron dynamics can be reduced to a one-dimensional effective stochastic process for a representative single synaptic weight,
with a friction and noise terms that must be determined self-consistently: we call this a DMFE, 
in explicit analogy with the dynamical mean-field theory (DMFT) of strongly correlated electrons~\cite{GKKR96}.
Solving the DMFE cannot \textit{a priori} be performed
analytically even in very simple cases: it requires a numerical analysis which, as discussed in the conclusions (Sec.~\ref{sec-conclusion}), 
we leave for future work.

\section{The continuous random perceptron model and its dynamics}
\label{sec-def-model}

The random perceptron model that we study in this work is defined in the following way.
Consider a $N$-dimensional vector ${\vec X = \{x_i\}_{i=1\cdots N}}$ constrained to live on the $N$ dimensional hypersphere such that ${\vec X\cdot \vec X = \sum_{i=1}^{N}x_i^2= N}$, where the dot sign indicates the scalar product in $\mathbb{R}^N$.
The vector $\vec X$ has the interpretation of the set of synaptic weights $x_i$ --~they are the degrees of freedom of the perceptron~-- and in what follows we will focus on the `thermodynamic' limit of large $N$.

The Hamiltonian of the model (also called `loss function' in machine learning) is
\beq
 \mathcal{H} \left(\vec X\right) =\sum_{\mu=1}^M v(h_\mu)
 \ , \qquad
 h_\mu = r_\mu - w  \ ,
 \qquad
 r_\mu=\vec F^\mu\cdot  \vec X \:.
\label{eq-def-Hamiltonian-perceptron}
\eeq
The vectors $\vec F^\m$ have components $F^\mu_i$ which are independent Gaussian variables of zero mean and variance $1/N$.
The index $\mu$ runs from 1 to ${M = \a N}$, the ratio ${\a =M/N}$ being thus fixed in the limit of large $N$.
Therefore each vector $\vec F^\mu$ represents a random pattern of quenched disorder, to which the system has to accomodate, \textit{i.e.} a {\it constraint}.
In the following, the statistical average over this disorder will be denoted with an overbar, as for instance in ${\overline{F^\mu_i F^\nu_j} = \frac{1}{N} \delta_{ij} \delta_{\mu \nu}}$.

The dynamical mean-field equations can be derived for arbitrary potentials $v(h)$.
In the case of the continuous perceptron, and for models related to hard-spheres, a particular relevant choice corresponds to ${v(h)=0}$ for ${h>0}$ and ${v(h)>0}$ for ${h<0}$.
In this way, a configuration that satisfies all the constraints ${h_\mu>0}$ has zero energy, and a configuration for which there is at least one unsatisfied constraint has positive energy.
Thus, working at zero temperature one can select configurations violating the minimum number of constraints. 
A typical choice is the harmonic soft potential for which ${v(h)=h^2 \theta(-h)/2}$ with $\theta(h)$ the Heaviside step function.
Finally, ${w \in \mathbb{R}}$ is a free parameter of the model.
It can be used to tune the degree of non-convexity of the model, once seen from a constrained optimization point of view~\cite{FPSUZ17}; the non-convex regime corresponds to ${w<0}$.

Besides the machine-learning perspective, one can think of this model as describing a single point particle of coordinate $\vec X$ moving on the surface of the $N$-dimensional sphere of radius $\sqrt{N}$. On the sphere, there are $M$ randomly distributed quenched obstacles with coordinates ${\vec Y^\m = -{\sqrt{N} \vec F^\m}}$ (in such a way that $\overline{\vec Y^\m \cdot \vec Y^\m} = N$, so that the vectors $\vec Y^\mu$ are on the surface of the sphere).
The distance between the particle and an obstacle is, when ${h_\mu > 0}$:
\beq
 h_\mu > 0 
 \qquad \Leftrightarrow \qquad
 | \vec X - \vec Y^\m |^2 \sim 2N - 2 \vec X \cdot \vec Y^\mu
 = 2 N + 2 \sqrt{N} \vec X \cdot \vec F^\m
 > 2 N + 2\sqrt{N} w
 \ ,
\eeq
where, therefore, ${2N + 2\sqrt{N} w}$ is the radius of an obstacle.
In analogy with soft sphere models of glasses, we say that the particle  overlap with the obstacle $\m$ whenever ${h_\mu<0}$ and therefore we call $h_\mu$ a gap variable.

The Langevin dynamics of this model is defined by a set of coupled stochastic equations for the synaptic weight $x_i(t)$ at time $t$, with ${i=1, \dots, N}$:
\beq
m \ddot x_i(t) +
 \int_{0}^t \de t' \, \G_R(t,t') \, \dot x_i(t')
 = -\hat \n(t) x_i(t) - \frac{\partial \mathcal{H}(\vec X(t))}{\partial x_i(t)} +  \h_i(t) \ ,
 \qquad
 \la \h_i(t) \h_j(t') \ra
 =  \d_{ij} \G_C(t,t')
 \ ,
\label{eq:Lang}
\eeq
where
$\dot x_i$ denotes as usual a time derivative,
$\hat \n(t)$ is a Lagrange multiplier that is used to impose the spherical constraint ${|\vec X(t)|^2=N}$ at each time $t$, and is usually self-consistently computed at the end of the computation.
$\h_i(t)$ is a Gaussian colored noise of zero mean, and the brackets denote the statistical average over this noise.
The kernel $\G_R(t,t')$ represents a friction term while $\G_C(t,t')$ provides the covariance function of a general correlated Gaussian noise; at equilibrium they are related by the fluctuation-dissipation theorem, but here we keep them generic in order to include the out-of-equilibrium cases.
The inertial term $m\ddot x_i(t)$, which is introduced in Eq.~\eqref{eq:Lang} for completeness, can be dropped in the so-called {\it overdamped limit} of the
Langevin equation, corresponding to $m\to 0$. In the following, to avoid keeping track of this term, we stick to the overdamped limit, but the inertial term can be reintroduced at any time in the derivation.

The stochastic initial configuration of $\vec X$ at time ${t=0}$ is extracted randomly from an arbitrary probability distribution which samples the whole phase space;
here we choose to consider the equilibrium distribution at inverse temperature $\b_g$:
\beq
\label{eq-pdf-initial-condition-equilibrium}
\begin{split}
 P_N(\vec X (0))
 	=\frac{1}{Z_N(\b_g)} \exp \argc{-\b_g\HH(\vec X (0))}
 \ , \qquad 
 Z_N(\b_g)
	&=\int_{\SSS_N} \de \vec X (0) \exp \argc{-\b_g\HH(\vec X (0)}
 \ ,
\end{split}
\eeq
where $Z(\b_g)$ is the partition function at inverse temperature $\b_g$ that corresponds to the integral over the sphere $\SSS_N$ defined by the spherical constraint.
We have thus three independent sources of stochasticity in the perceptron model, so we must consider three combined statistical averages:
first over the initial condition $\arga{x_i (0)}_{i=1,\dots,N}$ according to Eq.~\eqref{eq-pdf-initial-condition-equilibrium},
second over the colored noise $\arga{\eta_i(t)}_{i=1,\dots,N}$,
and third over the quenched disorder ${\arga{F_i^{\mu}}}_{i=1,\dots,N}^{\mu=1,\dots,M}$.
The corresponding averages will be denoted by ${\moy{\cdots}}$ for the combined noise and initial condition
--~otherwise it might be written explicitly as ${\moy{\cdots}_{\vec{X}(0)}}$ or ${\moy{\cdots}_{\eta}}$~--
and $\overline{\cdots}$ for the quenched disorder.

The resulting stochastic dynamics depends on the explicit form and properties of the two kernels $\G_R$ and $\G_C$.
Note that ${\G_R(t,t'>t)=0}$ by causality, while ${\G_C(t,t') = \G_C(t',t)}$ by definition.
We will only consider cases where these kernels are time-translational invariant, 
\textit{i.e.} $\G_R(t,t') = \G_R(t-t')$ and $\G_C(t,t') = \G_C(t-t')$, with $\G_R(t<0)=0$ and $\G_C(t) = \G_C(-t)$,
though we will keep the generic dependence on the two times as long as possible in our derivations.
Special cases are the following:
\begin{itemize}
\item For an equilibrium thermal bath, both $\G_C(t-t')$ and $\G_R(t-t')$ are time-translationally invariant. Moreover,
a fluctuation-dissipation-theorem (FDT) holds~\cite{ZBCK05}, in the form ${\G_C(t) = T [\G_R(t) + \G_R(-t) ]}$ or equivalently ${\G_R(t) = \b \th(t) \G_C(t)}$.
We emphasise that the inverse temperature of the thermal bath ${\b=1/T}$ can be different from the inverse temperature $\b_g$ of the distribution of the initial configuration.
If $\b=\b_g$, we are considering fully equilibrium dynamics. If instead ${\b \neq \b_g}$, we are considering a system prepared in equilibrium at $\b_g$ and then instantaneously quenched at a different temperature~$\b$.
\item
More specifically, the equilibrium case with white noise corresponds for instance to the choice ${\G_C(t) = T \g \frac{1}{\t} e^{-|t|/\t}}$, and then ${\G_R(t) = \g \frac{1}{\t} e^{-t/\t} \th(t)}$,
in the limit ${\t\to 0}$.
One recovers in this case a regular friction term ${\g \dot X_i(t)}$ and a noise kernel
${\G_C(t,t') = 2 T \g \d(t-t')}$.
\item
Standard active matter models correspond to ${\G_R(t) = \g \frac{1}{\t} e^{-t/\t} \th(t)}$ for ${\t\to0}$, hence a regular friction term, with a noise kernel
${\G_C(t) = 2 T \g \d(t-t') + \G_a(t)}$, where $\G_a(t)$ describes the active part.
In this case FDT is violated by construction, and the term $\G_a(t)$ can be interpreted as coming from particles self-propulsion~\cite{BK13}.
\item
A force constant in time, and random Gaussian for each component $i$, is described by the same structure as above but with ${\G_a(t) = f_0^2}$.
This corresponds to a random constant drive of the system.
\end{itemize}
We emphasise that we will consider generic kernels $\Gamma_R$ and $\Gamma_C$ in all the derivations that follow.
We will only need to distinguish between their regular and singular parts when deriving the dynamics from the supersymmetric path-integral formulation in Sec.~\ref{sec-SUSY}.

Given this definition of the continuous random perceptron model,
our goal will be to determine which is the stochastic process that would effectively describe the large-$N$ fluctuations of the gaps ${h_{\mu}(t)}$, or rather of ${r_{\mu}(t)=h_{\mu}(t) + w}$,
as a combination of the three sources of stochasticity present in the model in the thermodynamic limit.
In other words, we will derive the mean-field dynamics of a single variable, which could be either one typical synaptic weight ${x(t)}$, one typical gap ${h(t)}$, or one typical reduced gap ${r(t)=h(t)+w}$.
The corresponding statistical average will be denoted respectively by ${\moy{\cdot}_{x}}$,${\moy{\cdot}_{h}}$ or ${\moy{\cdot}_{r}}$, including both the average over the different histories of the effective stochastic process and over its stochastic initial condition.
We emphasise that $N$ is simultaneously the number of degrees of freedom, the radius of the hypersphere, and the fluctuations of disorder; this thermodynamic limit is thus a specific joint limit of these three quantities for the model.

In Sec.~\ref{sec-cavity-method} we present the derivation based on the cavity method, postponing to Sec.~\ref{sec-SUSY} its complementary counterpart based on the supersymmetric path-integral.

\section{Derivation of the dynamical mean field equations through the cavity method}
\label{sec-cavity-method}
%

The random perceptron defined in Sec.~\ref{sec-def-model} is a fully-connected model, 
because the Hamiltonian~\eqref{eq-def-Hamiltonian-perceptron} depends only on collective variables ${h_\mu(t)=\vec{F}^{\mu} \cdot \vec{X}(t)} - w$ which, given that the $F_i^\m$ are all non-zero and of the same order,
are averages over all the variables of the model.
The large-$N$ dynamics of such fully-connected models can usually be solved using the \emph{dynamical} cavity method, that is well described in Ref.~\cite{MPV87}.
The general idea can be summarised as follows.
Consider a dynamical system of $N$ fully-connected variables ${\lbrace x_i(t) \rbrace}$
--~such as the time-dependent synaptic weights of the perceptron~--
with ${i=1,\dots,N}$.
\begin{enumerate}
\item Write down the coupled dynamics and initial condition of the $N$ variables, which actually define the model under consideration.
\item Add a new dynamical variable ${x_0(t)}$, coupled in the same way to the initial $N$ variables.
The new variable will affect both the initial condition of the dynamics and the dynamical evolution itself.
\item Because the model is fully-connected, the coupling between this new variable and all the others is small in the thermodynamic limit ${N \to \infty}$, and it can thus be treated by perturbation theory.
\item Treating perturbatively the effect of the new variable both on the initial condition and on the dynamics, one can write a self-consistent effective dynamical process for the new variable.
\item Note that, after the new variable $x_0(t)$ is added, the system contains ${N+1}$ perfectly equivalent variables.
Hence, $x_0(t)$ has nothing special, and we could have repeated the same argument by choosing another variable instead. As a consequence,
the effective dynamical process for ${x_0(t)}$ can be promoted to the effective process characterising a typical variable ${x(t)}$, \textit{i.e.}~to the mean-field dynamics of the model.
\end{enumerate}
This strategy is the dynamical counterpart of a more standard and widespread \emph{static} cavity method, that was introduced in the context of spin glasses \cite{MPV87} and then fruitfully applied to constraint satisfaction and inference problems \cite{MPV86, MP01, MM09, KZ16}.
It relies on the fact that the large-$N$ limit of fully-connected models can be described by self-consistent mean-field equations.

Thereafter, given some preliminary definitions (Sec.~\ref{sec:lin-resp}),
we first use the dynamical cavity method to derive the dynamics of the effective stochastic process (Sec.~\ref{sec-cavity-method-derivation}),
and secondly we adapt a static cavity method to determine its corresponding initial condition (Sec.~\ref{sec-cavity-method-initial-condition}).
In both steps, we use the fact that in the large-$N$ limit, having ${N \pm 1}$ variables and ${M \pm 1}$ constraints (with ${M=\alpha N}$) can be treated self-consistently by perturbation theory.
These findings are summarised in Sec.~\ref{sec-cavity-method-summary}, providing a shortcut for the reader.
The corresponding evolution equations for the correlation and response functions are derived in Sec.~\ref{sec-cavity-method-correlation-reponse}.
Finally, we discuss the specific case of equilibrium, where the effective single-variable stochastic process simplies further thanks to the FDT relation, in Sec.~\ref{sec-cavity-method-equilibrium}.

\subsection{Perturbations and linear response}
\label{sec:lin-resp}

We consider a system with $N$ variables ${x_i(t)}$ (${i=1,\dots,N}$).
Using the explicit form of the Hamiltonian \eqref{eq-def-Hamiltonian-perceptron},
we start by rewriting their corresponding Langevin equations~\eqref{eq:Lang}:
\beq
 \frac{\partial \mathcal{H}(\vec{X}(t)) }{\partial x_i(t)}
 = \sum_{\mu=1}^{M} F^\mu_i v'\left( h_\mu (t) \right) 
 \quad \Rightarrow \quad
 \int_{0}^t \de t' \, \G_R(t,t') \, \dot x_i(t')
 = -\hat\nu(t) x_i(t) + \eta_i(t) - \sum_{\mu=1}^{M} F^\mu_i \, v'\left(h_\mu(t)\right)
 \:.
\label{eq-Langevin-explicit-cavity}
\eeq
As a preliminary step, we discuss two type of dynamic or static external fields that can be added to the perceptron model.
Here and in the following we consider these fields as infinitesimal, \textit{i.e.}~we focus on the linear response regime.
First, one can add a magnetic field $H_i$ on variable $i$. In the static version, this corresponds to changing the Hamiltonian as follows:
\beq
 \HH(\vec X) \to \HH(\vec X) - \sum_{i=1}^N H_i x_i
 \ .
\label{eq:staticHdef}
\eeq
In the dynamic version, it corresponds to changing the Langevin equation \eqref{eq-Langevin-explicit-cavity} as follows:
\beq
 \int_{0}^t \de t' \, \G_R(t,t') \, \dot x_i(t')
 = -\hat\nu(t) x_i(t) + \eta_i(t) - \sum_{\mu=1}^{M} F^\mu_i \, v'\left(h_\mu(t)\right) + H_i(t)
 \ .
\eeq
A different external field in which we will be interested amounts to adding a shift to a gap $\mu$, corresponding to ${h_\m \to h_\m + P_\m}$.
In the static version, this amounts to replacing
\beq
 \mathcal{H}(\vec{X}) \to 
 \sum_{\mu=1}^{M} v\left(h_\mu + P_\mu \right) \simeq \HH(\vec X) +
 \sum_{\mu=1}^M v'\left(h_\mu \right) P_\mu
 \ .
 \label{eq-Pdef}
\eeq 
In the dynamic version, the Langevin equation \eqref{eq-Langevin-explicit-cavity} is modified as follows:
\beq
\begin{split}
 \int_{0}^t \de t' \, \G_R(t,t') \, \dot x_i(t')
 &= -\hat\nu(t) x_i(t) + \eta_i(t) - \sum_{\mu=1}^{M} F^\mu_i \, v'\left(h_\mu(t) + P_\m(t) \right)
 \\
 & \simeq -\hat\nu(t) x_i(t) + \eta_i(t) - \sum_{\mu=1}^{M} F^\mu_i \, v'\left(h_\mu(t) \right) - \sum_{\mu=1}^{M} F^\mu_i \, v''\left(h_\mu(t) \right) P_\m(t) 
 \ .
\end{split}
\label{eq-Pdefdyn}
\eeq
Note that adding a single dynamic field $P_\m(t)$ on a given gap $\m$ is equivalent to adding a field
${H_i(t) = - F^\m_i \, v''\left(h_\mu(t) \right) P_\m(t)}$ on each of the variables ${i = 1, \dots, N}$.
This leads to the following useful identity between dynamical responses: 
for any observable $\OO(t)$, one has
\beq
\begin{split}
 \d \OO(t)
 &= \int_0^t \de t' \frac{\d \OO(t)}{\d P_\m(t')} P_\m(t')
	= -  \sum_{i=1}^N \int_0^t \de t' \frac{\d \OO(t)}{\d H_i(t')} F^\m_i \, v''\left(h_\mu(t') \right) P_\m(t)  \\
 & \Rightarrow \qquad
 \frac{\d \OO(t)}{\d P_\m(t')}
 	= -  \sum_{i=1}^N \frac{\d \OO(t)}{\d H_i(t')} F^\m_i \, v''\left(h_\mu(t') \right) 
 \ .
\end{split}
\eeq

\subsection{Derivation of the self-consistent process for the dynamics: the dynamical cavity method}
\label{sec-cavity-method-derivation}

To the system of $N$ variables ${i=1,\dots, N}$,
we now add a new dynamical variable $x_0(t)$, along with $M$ new quenched variables $F_{0}^\mu$ (${\mu=1,\dots,M}$) and the additional noise ${\eta_0(t)}$, generalising the model to ${N+1}$ degrees of freedom.
We emphasise that from now on, we will denote ${x_i(t)}$ only the variables for ${i>0}$ and treat $x_0(t)$ separately, except if indicated otherwise.
The updated gap variables $h_\mu(t)$, which drive the dynamics via the potential $v(h)$, can then be decomposed as
\beq
 h_\mu (t) = h_\mu^{(N)}(t) + P_\mu(t)
 \ , \quad \quad \text{with} \quad h_\mu^{(N)} (t)
 = \sum_{i=1}^{N} F_i^\mu x_i(t) -w \: \,\text{and}\:\,P_\mu(t)=F^{\mu}_{0}x_{0} (t)
 \,.
\label{eq:hnuexp}
\eeq
From the point of view of the original $N$ variables, therefore, the introduction of the new variable $x_0(t)$ corresponds to a perturbation ${P_\mu(t) =F^{\mu}_{0}x_{0} (t)}$ on the gap $h_\mu$.
We emphasise that by definition the gaps ${h_\mu^{(N)}(t)}$ are thus unperturbed variables uncorrelated to ${F_{0}^\mu}$.
Because the quenched disorder fluctuates in distribution according to ${(F_{0}^\mu)^2 \simeq 1/N}$,
and the variable $x_0$ is itself of order 1 when ${N \to \io}$, one has ${P_\m \propto 1/\sqrt N}$,
so that we can treat it perturbatively in linear response.
Note that the Lagrange multiplier ${\hat\nu(t)}$ now enforces the spherical constraint on the ${N+1}$ components of ${\vec{X}(t)}$ as ${\vec{X}(t)^2=N+1}$, but it will be self-consistently fixed at the end of the computation, so for the time being we leave it as a time-dependent unknown parameter.

The perturbation $P_\m$ appears both as a dynamic perturbation in the Langevin equation, and as a static perturbation in the initial condition.
Therefore, one can write in perturbation theory at large $N$ that
\beq
 x_i(t)
 = x_i^{(0)}(t) + \d x_i^{\textrm{(dyn)}}(t) + \d x_i^{\textrm{(in)}}(t)
 \ ,
\label{eq:deltax2}
\eeq
where ${x_i^{(0)}(t)}$ is the solution of the Langevin equations of the $N$ variables in absence of any perturbation,
with a stochastic initial condition distributed according to Eq.~\eqref{eq-pdf-initial-condition-equilibrium}, and
the large-$N$ perturbative correction to ${x_i^{(0)}(t)}$ can be decomposed in two parts.
First, ${\delta x_i^{(\textrm{dyn})}(t)}$ is produced by the fact that the new variable $x_0$ perturbs the dynamical equation through the external field $P_\m(t)$.
Secondly, ${\delta x_i^{(\textrm{in})}(t)}$ follows from the fact that the initial condition for the variables $x_i$ 
is also changed by the presence of the variable $x_0$, trough a static field $P_\m$.
Formally, these perturbations can be written using functional derivatives:
\begin{eqnarray}
 \delta x_i^{(\textrm{dyn})}(t)
 = \sum_{\mu=1}^{M}\int_{0}^{t} \de t' \, P_\m(t') \, \left[ \frac{\delta}{\delta P_\m(t')} x_i(t) \right\vert_{\scriptsize{\begin{array}{l} P_\m(t)=0 \end{array}}}
 &=&\sum_{\mu=1}^{M} \int_{0}^{t} \de t' \, \frac{\delta x_i^{(0)}(t) }{\delta P_\m(t')} \, P_\m(t')
 \ ,
 \label{eq-delta-x-dyn-formal}
 \\
 \delta x_i^{(\textrm{in})}(t)
 = \sum_{\mu=1}^{M} P_\mu(0) \, \left[ \frac{\delta}{\delta P_\mu(0)} x_i(t) \right\vert_{\scriptsize{\begin{array}{l} P_\mu(t)=0 \end{array}}}
 &=& \sum_{\mu=1}^{M}  \, \frac{\delta x_i^{(0)}(t)}{\delta P_\mu(0)}\, P_\mu(0)
 \ ,
\label{eq-delta-x-in-formal}
\end{eqnarray}
where we have introduced the slightly abusive notations $\frac{\delta x_i^{(0)}(t)}{\delta P_\m(t')}$ and $\frac{\delta x_i^{(0)}(t)}{\delta P_\mu(0)}$ 
to denote the functional derivatives evaluated at ${P_\mu(t)=0}$, \textit{i.e.}~on the unperturbed dynamics for the $\{x_i(t)\}$.
Note that ${\delta x_i^{\textrm{(dyn)}}(0)=0}$ because of its time-integral,
and thus at time ${t=0}$ the only corrections to $x_i(t)$ come from the modification of the Boltzmann measure,
as it should be.
We emphasise that in all the derivations that follows, there will always be \textit{a priori} two such corrections to take into account, first in the dynamics itself and secondly propagating from the initial condition.

Let us now consider the dynamical equation for the new variable $x_{0}$ itself, and apply a similar perturbative treatment.
The variable $x_0$ follows the same full Langevin dynamics, Eq.~\eqref{eq-Langevin-explicit-cavity}, as the others variables $x_i$, hence we have that
\beq
\begin{split}
\int_{0}^t \de t' \, \G_R(t,t') \, \dot x_0(t')
 &\stackrel{\eqref{eq-Langevin-explicit-cavity}}{=}
 	-\hat \nu(t) x_{0}(t) + \eta_{0}(t)
 		-\sum_{\mu=1}^M F^{\mu}_{0} v'\left(h_\mu(t)\right)
 \\
 & \stackrel{\eqref{eq:hnuexp}}{\simeq}
 	-\hat \nu(t) x_{0}(t) + \eta_{0}(t)
 		-\sum_{\mu=1}^M F^{\mu}_{0} v'\left(h_\mu^{(N)}(t)\right)
 	- \sum_{\mu=1}^M \left(F^{\mu}_{0}\right)^2 v''\left(h_\mu^{(N)}(t)\right) \, x_0(t)
 \ .
\end{split}
\label{cavity0}
\eeq
Since the $(F^{\mu}_0)^2$ are i.i.d. variables with average equal to $1/N$ the third term in Eq.~\eqref{cavity0} radically simplifies in the thermodynamic limit ${N \to \infty}$ (and hence ${M=\alpha N \to \infty}$). Its distribution concentrates around its average value up to fluctuations of the order $1/\sqrt{N}$:
\beq
 \sum_{\mu=1}^M \left(F^{\mu}_{0}\right)^2 v''\left(h_\mu^{(N)}(t)\right) \, x_0(t)
 \stackrel{(N \to \infty)}{\to}
 \a \overline{\moy{v''(h(t))}}_h  \, x_0(t)
 \: ,
\eeq
where $\overline{\moy{\cdots}}_h$ is the first occurrence of the statistical average over the effective stochastic process, that we are aiming to characterise, for the typical gap $h(t)$.
Note that also in the case of the effective process the average is separated into an average over thermal noise and initial condition, and an average over disorder.
The third term in Eq.~\eqref{cavity0} can thus be integrated into the Lagrange multiplier, which gives
\beq\label{eq-cavity1-redef-Lagrange}
\begin{split}
 \int_{0}^t \de t' \, \G_R(t,t') \dot x_0(t')
 &= -\tilde \nu(t) x_{0}(t) + \eta_{0}(t)
 		-\sum_{\mu=1}^M F^{\mu}_{0} v'\left(h_\mu^{(N)}(t)\right)
 \ , \\
 \tilde \n(t)&=\hat\nu(t) + \a \overline{\moy{v''(h(t))}}_h
 \ .
\end{split}
\eeq
The next step is to expand the gaps $h_\mu^{(N)}(t)$ appearing in Eq.~\eqref{eq-cavity1-redef-Lagrange} by making use of the expansion in Eq.~\eqref{eq:deltax2}.
We introduce the unperturbed gaps $h_\mu^{(0)}(t)$, that depend only on the $N$ original spins and evolve in absence of the new spin, by
\beq
 h_\mu^{(0)}(t)  = \sum_{i=1}^N F^\mu_i x_i^{(0)} (t) -w 
 \qquad
 \Rightarrow
 \qquad
  h_\mu^{(N)}(t) =  h_\mu^{(0)}(t) + \d h_\mu^{\textrm{(dyn)}}(t) +  \d h_\mu^{\textrm{(in)}}(t) \ ,
\eeq
where the perturbations have the same structure as in Eqs.~\eqref{eq-delta-x-dyn-formal}-\eqref{eq-delta-x-in-formal},
for instance
\beq
 \d h_\mu^{\textrm{(dyn)}}(t) = \sum_{i=1}^N F^\mu_i \d x_i^{\textrm{(dyn)}}(t) 
 = \sum_{\nu=1}^{M} \int_{0}^{t} \de t' \, \frac{\delta h_\mu^{(0)}(t) }{\delta P_\n(t')} \, P_\n(t')
 \ .
\eeq
Similarly, we can write
\beq
 v'\left(h_\mu^{(N)}(t)\right)
 = v'\left(h_\mu^{(0)}(t)\right)
	+\sum_{\nu=1}^{M} \frac{\delta v'\left(h_\mu^{(0)}(t)\right)}{\delta P_\nu(0)}P_\nu(0)
	+\sum_{\nu=1}^M  \int_{0}^t \de t' \frac{\delta v'\left(h_\mu^{(0)}(t)\right)}{\delta P_\n(t')} P_\n(t') \ .
\eeq
Plugging this expansion into Eq.~\eqref{eq-cavity1-redef-Lagrange}, and recalling that ${P_\n = F^\n_0 x_0}$,
we obtain
\beq
\begin{split}
\int_{0}^t \de t' \, \G_R(t,t') \, \dot x_0(t')
 \simeq
 & - \tilde \n(t) x_{0}(t) +\eta_{0}(t)
 	\underbrace{- \sum_{\mu=1}^M F^{\mu}_{0} v'\left(h_\mu^{(0)}(t)\right)}_{\text{(I)}}
 	\underbrace{ - \sum_{\mu=1}^M F^{\mu}_{0} \sum_{\nu=1}^{M} \frac{\delta v'\left(h_\mu^{(0)}(t)\right)}{\delta P_\nu(0)}F^\nu_0x_{0}(0)}_{\text{(II)}}
 \\
 & \underbrace{ -\sum_{\mu=1}^{M} F^\mu_{0} \sum_{\n=1}^M \int_{0}^t \de t'  \frac{\delta v'\left(h_\mu^{(0)}(t)\right)}{\delta P_\n(t')}  F^\nu_{0}
 x_{0}(t')}_{\text{(III)}} \ .
\end{split}
\label{cavity1}
\eeq
We will now successively examine the contributions (I)-(II)-(III) in the thermodynamic limit; to guide the intuition regarding the effective stochastic process we are aiming at, (I) will correspond to its effective noise, (II) to its correlation memory kernel, and (III) to its response memory kernel.
We start by (III). In order to evaluate this term one has to take into account that 
$\frac{\delta v'\left(h_\mu^{(0)}(t)\right)}{\delta P_\n(t')}$
in fully connected models, the response of a variable $\m$ to a field on variable $\n$
is generically of the order one for ${\mu=\nu}$ and of the order $1/N$ for ${\mu \ne \nu}$
(this is because responses are related to correlations, and correlations of distinct variable vanish as $1/N$ in fully connected models).
In consequence, in the large-$N$ limit, only the diagonal terms $\m=\n$ in (III) survive, and
the distribution of (III) concentrates around its average up to fluctuations of the order $1/N$ (as it can be checked computing its variance): 
\beq
\begin{split}
 \text{(III)}
 \stackrel{(N \to \infty)}{\to}
 \int_{0}^t \de t' M_R(t,t') \, x_{0}(t')
 \ , \quad \quad  \text{with} \;\;
 M_R(t,t') \equiv - \alpha \, \frac{\delta \overline{\langle v'(h(t)) \rangle}_h }{\delta P(t')}
 \ ,
\end{split}
\label{eq-cavity1-III}
\eeq
where ${M_R(t,t')}$ is the two-time response memory kernel.
Here the average must be intended as the average over the dynamical history of the variables $\{x_i\}_{i=1,\ldots, N}$ 
when the variable $x_0$ is completely absent, and it is then replaced by an average over the effective process -- that we still have to characterise.
Following the same reasoning, the term (II) in Eq.~\eqref{cavity1} simplifies in the large $N$ limit:
\beq
 \text{(II)}
 \stackrel{(N \to \infty)}{\to} \b_g M_{C}^{(c)}(t,0) \, x_0(0)
 \label{eq-cavity1-II}
\eeq
where, by taking the derivative over the static field $P_\mu (0)$ associated with the initial condition, Eq.~\eqref{eq-pdf-initial-condition-equilibrium}, 
one obtains the connected correlation memory kernel ${M_C^{(c)}(t,t')}$, defined
as the thermodynamic limit of the connected two-time second cumulant of the force $v'$:
\beq
\begin{split}
 \frac{1}N \sum_{\mu=1}^{M}\left[
 	\overline{\moy{ v'(h_\mu^{(0)}(t)) v'(h_\mu^{(0)}(t'))}}
 	- \overline{\moy{ v'(h_\mu^{(0)}(t))} \moy{ v'(h_\mu^{(0)}(t'))}}
 \right] &
 \\
 \stackrel{(N \to \infty)}{\to}
 \alpha \argc{ \overline{\moy{ v'(h(t))v'(h(t'))}}_h - \overline{\moy{v'(h(t))}_h \moy{v'(h(t'))}}_h }
  & \equiv M_C^{(c)}(t,t')
\end{split}
\label{eq-cavity1-II-bis}
\eeq
that once again should in the end be computed as an average over the effective process.
Finally, the term (I) in Eq.~\eqref{cavity1} can be combined with the original noise ${\eta_0(t)}$ in order to define the effective noise
\beq
 \Xi(t)
 = \eta_{0}(t)-\sum_{\mu=1}^M F^{\mu}_{0} v'\left(h_\mu^{(0)}(t)\right)
 \:.
\label{eq-cavity1-I}
\eeq
Combining Eqs.~\eqref{eq-cavity1-redef-Lagrange}, \eqref{eq-cavity1-III}, \eqref{eq-cavity1-II} and \eqref{eq-cavity1-I}, the large-$N$ Langevin equation for the new variable ${x_0(t)}$ can be rewritten as
\beq
\begin{split}
\int_{0}^t \de t' \, \G_R(t,t') \dot x_0(t')
 &= - \tilde \n(t) x_0(t) +\Xi(t)
 	+ \b_g M_{C}^{(c)}(t,0)x_0(0)
 	+ \int_{0}^t \de t' M_R(t,t') x_0(t')
 \ .
\end{split}
\label{effective_spin}
\eeq
To conclude the derivation, we need to specify the correlation function of the effective noise $\Xi(t)$.
First, the dynamics of $x_i^{(0)}(t)$ is independent of the additional noise $\eta_0(t)$ and therefore the two contributions to $\Xi(t)$ are statistically independent.
Then, one can split $\Xi(t)$ in fluctuations due to quenched disorder and due to noise and initial conditions. Indeed,  let us decompose $\Xi(t)$, for a given quenched disorder, into its average ${\moy{\cdots}}$ (over the noise and initial condition) and a fluctuating part:
\beq
\begin{split}
 \Xi(t)
 &= \eta_0(t)-\sum_{\mu=1}^M F^{\mu}_{0} v'\left(h_\mu^{(0)}(t)\right) 
 = \z_d(t) + \z(t) \ ,
 \\
 \text{with} & \quad\quad \z_d(t)
 	=-\sum_{\mu=1}^M F^{\mu}_{0} \left\langle v'\left(h_\mu^{(0)}(t)\right)\right\rangle \ ,
 \\
 \text{and} & \quad\quad \z(t)
 	=\eta_0(t) - \sum_{\mu=1}^M F^{\mu}_{0} \argp{v'\left(h_\mu^{(0)}(t)\right) - \moy{v'\left(h_\mu^{(0)}(t)\right)} } \ .
\label{noises}
\end{split}
\eeq
The fluctuations of the term $\z_d(t)$ are only due to the quenched disorder, and in the thermodynamic limit it becomes a Gaussian variable due to the central limit theorem.
Instead, $\z(t)$ primarily fluctuates due to the thermal noise and the initial condition;
it also becomes a Gaussian variable due to the central limit theorem.
Its fluctuations over the disorder concentrate in the thermodynamic limit and can be neglected.
One has therefore
\begin{eqnarray}
 & \overline{\z_d (t)} =0 \ , \quad
 & \overline{ \z_d (t)\z_d(t')}
 	= \a \overline{ \moy{v'(h(t))}_h \moy{ v'(h(t'))} }_h
 	\equiv M_d(t,t') \ ,
 \label{noises-corr-zetad}
 \\
 & \moy{ \zeta (t)} =0 \ , \quad
 & 
 \moy{ \z(t)\z(t')}
 	= \G_C(t,t') + M_C^{(c)} (t,t') \: , 
  \label{noises-corr-zeta}
\end{eqnarray}
with ${\Gamma_C}(t,t')$ the noise kernel introduced in Eq.~\eqref{eq:Lang}.
The noises $\z_d(t)$ and $\z(t)$ can therefore be thought as two independent noises, the first representing the fluctuations of the disorder, the second representing the thermal fluctuations.

The effective equation \eqref{effective_spin} for ${x_0(t)}$ is representative of the dynamics of all variables in the thermodynamic limit of such fully-connected models, leading to an exact mean-field description.
We can thus promote the effective dynamical process for ${x_0(t)}$ of Eq.~\eqref{effective_spin} to the effective process characterising a typical variable ${x(t)}$:
\beq
\begin{split}
 \int_{0}^t \de t' \, \G_R(t,t') \, \dot x(t')
 &= - \tilde \n(t) x(t) + \b_g M_{C}^{(c)}(t,0) \, x(0)
 		+  \int_{0}^t \de t' M_R(t,t') \, x(t')
 		+ \z_d(t) + \z(t)\:.
\end{split}
\label{effective_spin2}
\eeq
However, in order to close the equations, we have to write down the counterpart of Eq.~\eqref{effective_spin2} for the effective gap variable $h(t)$, or equivalently for the reduced gap ${r(t)=h(t)+w}$.
Indeed, the full dynamics depends on the gaps ${h_\mu(t)}$ rather than on the individual ${x_i(t)}$,
and the different functions defined in order to obtain the effective formulation
--~${\tilde{\nu}(t)}$, ${M_R(t,t')}$, ${M_C^{(c)}(t,t')}$, and ${M_d(t,t')}$~--
are in fact statistical averages of combinations of ${v'(h(t))}$ and ${v''(h(t))}$,
with ${h(t)}$ the typical gap in the thermodynamic limit.

We thus consider a system with $N$ variables and ${M=\alpha N}$ constraints labeled by ${\mu = 1, \dots, M}$, and we add a new constraint ${\vec{F}^{0}}$ that we will treat perturbatively in the large-$N$ limit, in the same spirit as before.
The corresponding new gap is ${h_{0}(t) = \sum_{i=1}^{N} F_i^{0} x_i(t) - w}$.
The effective dynamical equation for ${x_i(t)}$, in presence of the additional constraint, is still given by 
Eq.~\eqref{effective_spin2} (where an index $i$ is added to $x(t)$), with the additional term coming from the new constraint:
\beq
\begin{split}
 \int_{0}^t \de t' \, \G_R(t,t') \, \dot x_i(t')
 &= - \tilde \n(t) x_i(t) + \b_g M_{C}^{(c)}(t,0) \, x_i(0)
 		+  \int_{0}^t \de t' M_R(t,t') \, x_i(t')
 		+ \z^i_d(t) + \z^i(t) - F_i^{0} v'(h_{0}(t))
		\:.
\end{split}
\label{effective_spin3}
\eeq
Here, $\z^i$ and $\z_d^i$ are at leading order independent copies of the noise terms defined in Eqs.~\eqref{noises-corr-zetad}-\eqref{noises-corr-zeta}.
Note that the last term coming from the additional constraint is of order $1/\sqrt{N}$, and it vanishes
in the thermodynamic limit $N\to \io$ with $M=\a N$, giving back the effective equation~\eqref{effective_spin2} for the representative $x(t)$.
Adding one more constraint is a vanishing perturbation in the thermodynamic limit.
However, from the point of view of gaps, these $1/\sqrt{N}$ terms sum coherently and give a finite contribution.
Indeed, the self-consistency equation for the gap variable is obtained by multiplying Eq.~\eqref{effective_spin3}  by $ F^{0}_i$, and summing over $i$.
Denoting the reduced gap $r_{0}(t) = \vec{F}^{0} \cdot \vec{X}(t)=h_{0}(t)+w$, and dropping the index $0$ because all gaps are equivalent,
we obtain from Eq.~\eqref{effective_spin3} the following equation for the representative gap $r(t)$:
\beq
\int_{0}^t \de t' \, \G_R(t,t') \dot r(t')
 =  - \tilde \n(t) r(t)
 	+\b_g M_{C}^{(c)}(t,0) \, r(0)
 	+ \int_{0}^t \de t' M_R(t,t')r(t')
 	+ \z_d(t) + \z(t)
	- v'(r(t)-w)
 \label{eff_gap}
\eeq
The new noises ${\z(t) = \sum_{i=1}^N F_i^{0} \z^i(t)}$ and ${\z_d(t) = \sum_{i=1}^N F_i^{0} \z_d^i(t)}$ are linear combination of Gaussian variables, and therefore they are themselves Gaussian variables; thanks to the property ${\overline{ F^{0}_i F^{0}_j } = \d_{ij} /N}$, they have exactly the same statistics as $\z(t)$ and $\z_d(t)$ given in Eqs.~\eqref{noises-corr-zetad}-\eqref{noises-corr-zeta}.

We gather thereafter the different functions introduced along the derivation, 
that are then self-consistently defined with respect to the effective stochastic process for ${r(t)}$, or equivalently ${h(t)}$, given in Eq.~\eqref{eff_gap}:
\begin{eqnarray}
 \tilde{\nu}(t)
 & \stackrel{\eqref{eq-cavity1-redef-Lagrange}}{=} &
 	 \hat\nu(t) + \a \overline{\moy{v''(h(t))}}_h
 \ ,
 \\
 M_R(t,t')
 & \stackrel{\eqref{eq-cavity1-III}}{=} &
 	\alpha \, \frac{\delta \overline{ \langle v'(h(t)) \rangle}_h }{\delta P(t')}
 \ ,
 \\
 M_C^{(c)}(t,t')
 & \stackrel{\eqref{eq-cavity1-II-bis}}{=} &
	\alpha \argc{ \overline{\moy{ v'(h(t))v'(h(t'))}}_h - \overline{\moy{v'(h(t))}_h \moy{v'(h(t'))}}_h }
 \ ,
 \\
 M_d(t,t')
 & \stackrel{\eqref{noises-corr-zetad}}{=} &
 	\a \overline{\moy{v'(h(t))}_h \moy{ v'(h(t'))}}_h
 \ ,
\end{eqnarray}
with $\alpha$ the fixed ratio of constraints to degrees of freedom, and the field $P(t)$ has to be added to Eq.~\eqref{eff_gap}, as in Eq.~\eqref{eq-Pdefdyn}, by replacing ${v'(r(t)-w) \to v'(r(t) + P(t) -w)}$. The Lagrange multiplier ${\hat\nu(t)}$ is self-consistently determined by enforcing the spherical constraint, ${\vec{X}(t)^2=N}$, at the end of the calculation, implying for the individual $x_i(t)$ and in particular for its typical value that ${\overline{\moy{x(t)^2}}=1}$.
It is important to stress that in the effective process, the average $\langle \cdots \rangle$ has become an average over the thermal noise $\z(t)$, and over the initial condition.

In summary, the equations (\ref{effective_spin2},\ref{eff_gap}) define a stochastic process containing kernels that can be computed self-consistently as observables of the process.
This is the dynamical mean-field theory of the spherical random perceptron.

It remains, however, to define the initial conditions.
If one starts the dynamics from random initial conditions, which corresponds to fixing $\beta_g=0$, then 
both the variables $x_i(0)$ and the gaps $h_\mu(0)$ are i.i.d. Gaussian random variables with zero mean and unit variance and the dynamical mean-field equations substantially simplify since several terms containing $\beta_g$ drop out.
The situation is instead more involved in the case in which the initial conditions are drawn from a Boltzmann measure at finite temperature.
We shall consider it in the next section and then summarise all the results in the following one.

Finally, let us stress that starting from the self-consistent stochastic process one can also derive the evolution equations of the correlation and response functions. This will be done in Sec.~\ref{sec-cavity-method-correlation-reponse}.

\subsection{Initial condition for the self-consistent process: the static cavity method}
\label{sec-cavity-method-initial-condition}

We now discuss the case in which
the dynamics starts from an equilibrium distribution at $\b_g > 0$, and deduce the corresponding initial condition
for the effective process (\ref{effective_spin2},\ref{eff_gap}).
In order to obtain it we need to consider the probability distribution of the initial condition for all the variables,
given in the definition of our model,
and then produce the marginal probability distribution of the variable $x_0$.
This can be done again using the cavity method, in its static version.
In order to do that there are several possibilities, as in Refs.~\cite{AFP16, KZ16}.
Here we reproduce for completeness the static cavity route of Ref.~\cite{Me89}, 
because it follows the same logic of its dynamical counterpart used in Sec.~\ref{sec-cavity-method-derivation}.
In Sec.~\ref{sec-cavity-method-equilibrium} we will see that the dynamical derivation that we have outlined above is consistent with the static cavity approach for the initial condition, when we consider the equilibrium case for the dynamics.

Let us recall the Boltzmann measure for the initial configuration of the system, first given in Eq.~\eqref{eq-pdf-initial-condition-equilibrium}:
\beq
 P_{N,M}(\vec X(0))
 = \frac{1}{Z_{N,M}(\b_g)} \exp\left[-\frac{\b_g \l}{2}\sum_{i=1}^N x_i(0)^2 -\b_g\sum_{\mu=1}^M v(h_{\mu}(\vec X (0)))\right]
\label{eq-pdf-initial-condition-Boltzmann}
\eeq
where we have underlined that the measure contains $N$ variables $x_i$ and ${M=\alpha N}$ constraints ${\vec{F}^{\mu}}$ with $N$ components each,
and we added a Lagrange multiplier $\l$ in order to enforce the spherical constraint at initial time, \textit{i.e.}~${|\vec X(0)|^2 = N}$.
In general, depending on the potential $v(h)$ and the temperature $\b_g$, the model can be found in different 
phases~\cite{GD88,MPV87,MP01,MM09}, see e.g.~Ref.~\cite{FPSUZ17} for an explicit computation of the full phase diagram in the case of
$v(h) = h^2 \th(-h)/2$ and $\b_g=\io$. In the simplest case, corresponding to a ``paramagnetic'' or ``liquid'' phase,
in the large-$N$ limit the Boltzmann measure describes a single pure state.
In other cases, corresponding to different spin glass phases, there are many coexisting pure states~\cite{MPV87,MM09}.
In the following, for simplicity, we restrict to the case where there is a single pure state, usually called the ``replica symmetric'' case both in the replica
and cavity literature. The results of this section thus hold in cases where the dynamics starts within the liquid or paramagnetic phase.
The case where there are many coexisting pure states requires a more complicated treatment~\cite{MPV87,MP01,MM09}.
There is, however, a special case, usually called ``dynamical 1RSB phase''~\cite{MM09,FPSUZ17}, in which there are many coexisting pure states,
but the system remains globally paramagnetic; in this case, the thermodynamics is still described by the replica symmetric structure~\cite{CC05,KMRSZ07,KZ10b}, 
and our results thus hold. This implies that the solution discussed in the following can be used to study the dynamics starting inside one of these pure states,
as discussed in Ref.~\cite{BBM96} for the $p$-spin model.

Moreover, specifically throughout this section, we will simply denote the initial condition by ${\vec{X}=\lbrace x_i \rbrace}$, dropping the indication of time $t=0$ to simplify the notation;
the brackets will correspond to the statistical average with respect to the Boltzmann measure~\eqref{eq-pdf-initial-condition-Boltzmann} at fixed disorder,
and the overline to the average with respect to the constraints.

We first want to determine the distribution of a typical gap ${h}$ or reduced gap ${r=h+w}$ resulting from Eq.~\eqref{eq-pdf-initial-condition-Boltzmann}.
Let us consider a system with $N$ variables and $M$ constraints. We consider a new vector
$\vec F^0$, a corresponding gap ${h_0=\vec F^0\cdot \vec X -w}$ and a reduced gap ${r_0=\vec F^0\cdot \vec X}$. However, for the moment, {\it the new constraint is not added to the Hamiltonian}.
First, we note that the variable $r_0$ is the sum of $N$ random variables and due to the central limit theorem it becomes Gaussian when $N\to\io$.
Its statistical properties are thus given as follows, at fixed disorder and in the large-$N$ limit:
\begin{eqnarray}
 & \langle r_0\rangle &
	=  \vec F^0 \cdot \langle \vec X\rangle
	\equiv \omega \ ,
\label{PDF-r0-mean}
 \\
 & \langle r_0^2 \rangle_c &
 	= \sum_{i,j} F^0_i F^0_j \argc{ \langle x_i x_j \rangle - \langle x_i  \rangle \langle x_j \rangle}
	\stackrel{(N\to\infty)}{\simeq}
		\frac{1}{N} \argc{ \langle \vec{X}^2 \rangle - \langle \vec{X} \rangle^2}
 	\simeq 1-q_g
\label{PDF-r0-variance}
 \\
 & \text{with} & \quad
 	q_g \equiv \frac 1N \sum_{i=1}^N \langle x_i\rangle^2
 \:.
\label{PDF-r0-def-q}
\end{eqnarray}
The quantity $q_g$ is the {\it overlap}, a crucial quantity of spin glass theory~\cite{MPV87}. Here we added the suffix ``g'' to specify that it is computed at the inverse temperature $\b_g$.
Secondly, $\omega$ is a Gaussian random variable that fluctuates due to the disorder $\vec F^0$.
Its statistical properties are given by
\beq
 \overline{\omega}  = 0
 \, , \quad
 \overline{\omega^2} = q_g
 \ .
\eeq
It follows from Eqs.~\eqref{PDF-r0-mean}-\eqref{PDF-r0-variance} that the distribution of $r_0$, in the system with $N$ variables and $M$ constraints, is
\beq
 P_{N,M} (r_0|\omega)
  =\frac{1}{\sqrt{2\pi(1-q_g)}}\exp\left[-\frac{(r_0-\omega)^2}{2(1-q_g)} \right]
 \ , \qquad
 P(\omega)
  = \frac{1}{\sqrt{2\pi q_g}}e^{-\omega^2/(2q_g)}
 \ . 
\label{Pomega}
\eeq
{\it When we add the new constraint $\vec F^0$ to the Hamiltonian},
the distribution of $r_0$ in the system with $M+1$ constraints, conditioned to ${\omega=\langle r_0 \rangle}$, 
is modified to
\beq
 P_{N,M+1} (r_0|\omega)
\propto \exp\left[-\frac{(r_0-\omega)^2}{2(1-q_g)} - \b_gv(r_0-w)\right]
 \ ,
\label{Pr}
\eeq
where the additional factor ${\exp \argc{- \b_gv(r_0-w)}}$ is due to 
the modification of Eq.~\eqref{eq-pdf-initial-condition-Boltzmann} because of the new constraint.
Consequently, the initial condition for the effective process Eq.~\eqref{eff_gap} for the typical initial reduced gap $r(0)$ is extracted from 
\beq
 P(r(0) |\omega)
\propto \exp\left[-\frac{(r(0)-\omega)^2}{2(1-q_g)} - \b_gv(r(0)-w)\right]
 \ , \ \ \ \ \ \ \ \ \ \ \ \
 P(\omega) = \frac{1}{\sqrt{2\pi q_g}}e^{-\omega^2/(2q_g)}\:.
\label{eq-initial-condition-r0-with-q}
\eeq
However, we still need to determine self-consistently the parameter $q_g$ controlling these distributions.

To do this, let us add a new variable $x_0$ to the system.
The new Boltzmann measure over $x_0$ reads
\beq
 P_{N+1,M}(x_0)
 	= \frac{1}{Z_{N+1,M}(\b_g)} \int \left(\prod_{i=1}^N \de x_i\right)
 	\exp \arga{
  		- \frac{\b_g \l}{2} \sum_{i=1}^N x_i^2
 		- \frac{\b_g\l}{2}x_0^2
 		- \b_g\sum_{\mu=1}^M \argc{v(h_{\mu}^{(N)})+\delta v_\mu}
 	} \ ,
\eeq
where we have to retain only the two (small) leading terms of the perturbative expansion,
\beq
 \delta v_\mu
 = F_0^\mu v'(h_\mu^{(N)})x_0 + \frac{1}{2}(F_0^\mu)^2 v''(h_\mu^{(N)})x_0^2
 \ .
\eeq
%
We rewrite the previous expression as 
\beq
 P_{N+1,M}(x_0)
 		\propto e^{- \frac{\b_g\l}{2}x_0^2} \langle e^{-\beta_g  \sum_{\mu=1}^M[F_0^\mu v'(h_\mu^{(N)})x_0 + \frac{1}{2}(F_0^\mu)^2 v''(h_\mu^{(N)})x_0^2]}     \rangle_N
 \ ,
\eeq
where the average $\langle \cdot \rangle_N$ is over the system in absence of $x_0$.
Because the argument of the exponential is small we can perform an expansion in cumulants; only the first two have to be retained in the large-$N$ limit.
As before, one can recognise that some of the resulting contributions becomes non-fluctuating, \textit{i.e.} their distributions concentrate around a fixed value, in the large $N$ limit.
The final result is that 
\beq
 P_{N+1,M}(x_0)
  \propto  \,
 	 \exp\left[-\frac{\b_g\tilde \l}{2}x_0^2- A x_0  \right]
 \ ,
\label{PDF-new-variable-x0}
\eeq
where
\begin{eqnarray}
 \tilde \l
 &=&
	\l +\a \overline{\langle v''(r-w)\rangle}
		- \a \b_g \left[\overline{\langle \left(v'(r-w)\right)^2\rangle-\langle v'(r-w)\rangle^2} \right] \ ,
 \label{eq-tilde-lambda}
 \\
 A &=& \b_g \sum_{\mu=1}^M F_0^\mu \langle v'(r-w)\rangle
 \qquad \text{fluctuates over the disorder with} 
 \\
 \overline{A} & =& 0 \ ,
\qquad \qquad
 \overline{A^2}
	= \b_g^2 \sum_{\mu, \nu=1}^M F_0^\mu F_0^\nu \langle v'(r-w)\rangle^2
	\simeq \a \b_g^2 \overline{\langle v'(r-w)\rangle^2}
 \ .
\label{A}
\end{eqnarray}
The brackets are now thermal averages over the random variable $r$ extracted according to the probability distribution $P(r | \omega)$, 
while the overlines are the averages over the disorder represented by $\omega$ with the Gaussian measure $P(\omega)$, both given in Eq.~\eqref{eq-initial-condition-r0-with-q}.
The connection with the parameter $q_g$ in the distributions~\eqref{eq-initial-condition-r0-with-q}
is made by using the definition of $q_g$ in Eq.~\eqref{PDF-r0-def-q} and the spherical constraint:
\begin{eqnarray}
 \langle x_0\rangle
	= \frac{ \int_{-\infty}^\infty \de x_0  \exp\left[-\frac{\b_g\tilde \l}{2}x_0^2-A x_0  \right] x_0}{\int_{-\infty}^\infty \de x_0  \exp\left[-\frac{\b_g\tilde \l}{2}x_0^2-A x_0  \right]}
 	= - \frac{A}{\b_g\tilde \l}
 & \quad\quad\quad\quad \Rightarrow \quad &
 	q_g	\stackrel{\eqref{PDF-r0-def-q}}{=} \overline{\langle x_0\rangle^2}
 		= \frac{\overline{A^2}}{(\b_g\tilde \l)^2} \ ,
 \\
 \langle x_0^2\rangle
 	= \frac{ \int_{-\infty}^\infty \de x_0  \exp\left[-\frac{\b_g\tilde \l}{2}x_0^2-A x_0  \right] x_0^2}{\int_{-\infty}^\infty \de x_0  \exp\left[-\frac{\b_g\tilde \l}{2}x_0^2-A x_0  \right]}
 	= \frac{A^2 + \b_g\tilde \l}{(\b_g\tilde \l)^2}
 & \quad\quad\quad\quad \Rightarrow \quad &
 	1	= \overline{\langle x_0^2\rangle}
 		= \frac{\overline{A^2} + \b_g\tilde \l}{(\b_g\tilde \l)^2}
 \: .
\end{eqnarray}
It follows that
\begin{eqnarray}
 &\text{\textit{(i)}}& \quad\quad
  \b_g\tilde \l
 	= \frac{1}{1-q_g} \ ,
\label{eq-self-consistency-relations1}
 \\
& \text{\textit{(ii)}}& \quad\quad
  \frac{q_g}{(1-q_g)^2}
 	= \overline{A^2} = \a \b_g^2 \overline{\langle v'(r-w)\rangle^2} \ ,
\label{eq-self-consistency-relations2}
 \\
 &\text{\textit{(iii)}}& \quad\quad
  \b_g\l
 	= \frac{1-2q_g}{(1-q_g)^2} -\a\b_g\overline{\langle v''(r-w)\rangle} + \a \b_g^2\overline{\langle \left(v'(r-w)\right)^2\rangle} \ ,
\label{eq-self-consistency-relations3}
\end{eqnarray}
and in particular Eq.~\eqref{eq-self-consistency-relations2} provides the self-consistency equation for $q_g$ that we were seeking, writing explicitly the statistical averages according to Eq.~\eqref{eq-initial-condition-r0-with-q}:
\beq
 \frac{q_g}{(1-q_g)^2}
 	= \a \b_g^2 \int_{-\infty}^\infty \frac{\de \omega \, e^{-\omega^2/(2q_g)}}{\sqrt{2\pi q_g}
 		}
 		\left[ \frac{\int_{-\infty}^\infty \de r \, v'(r-w) \exp\left[-\frac{(r-\omega)^2}{2(1-q_g)} - \b_gv(r-w)\right]}{\int_{-\infty}^\infty \de r \exp\left[-\frac{(r-\omega)^2}{2(1-q_g)} - \b_gv(r-w)\right]}
 		\right]^2 \ .
\label{static_cavity}
\eeq
These equations coincide with what can be obtained through the replica method in the RS phase where a single pure state exists~\cite{FPSUZ17}.
The generalisation to a more complicated RSB structure requires the introduction of a hierarchy of fields that describe the fluctuations of quantities such as $\omega$ in the different
pure states~\cite{MPV87}.

From all this, it follows that the effective single variable $x$ is distributed at time $t=0$ according to the following probability distribution combining Eqs.~\eqref{PDF-new-variable-x0}, \eqref{A} and \eqref{eq-self-consistency-relations2}:
\beq
P(x(0)|A) \propto \exp\left[-\frac{\b_g\tilde \l}{2}x(0)^2-A x(0)  \right]
\ , \ \ \ \ \  \ \ \ \ \ \ \ P(A)=\sqrt{\frac{(1-q_g)^2}{2\pi q_g}}\exp\left[-\frac{(1-q_g)^2}{2q_g}A^2\right] \ ,
\label{eq-PDF-x0-A-static-cavity}
\eeq
where $q_g$ satisfies Eq.~\eqref{static_cavity}.
This concludes the characterisation of the probability distributions of the effective single variable and reduced gap at time $t=0$.
We will see that these probability distributions can be obtained also from the path integral approach in Sec.~\ref{sec-SUSY}.

\subsection{Summary: the single-variable effective stochastic process}
\label{sec-cavity-method-summary}

Here we summarise the effective processes that we have obtained in Sec.~\ref{sec-cavity-method-derivation} and Sec.~\ref{sec-cavity-method-initial-condition}.
We recall that ${\beta_g}$ is the inverse temperature of the Boltzman measure~\eqref{eq-pdf-initial-condition-equilibrium} chosen as initial condition, while $\Gamma_R$ and $\Gamma_C$ are respectively the friction and noise kernels defined and discussed in Sec.~\ref{sec-def-model}.
Our main results are:
\begin{enumerate}

\item
For the typical variable $x(t)$ the dynamics is given by Eqs.~\eqref{noises-corr-zetad}, \eqref{noises-corr-zeta} and \eqref{effective_spin2}:
\beq
\begin{split}
 & \int_{0}^t \de s \, \G_R(t,s) \, \dot x(s)
 =  - \tilde \n(t) x(t) + \b_g M_{C}^{(c)}(t,0) \, x(0)
 		+  \int_{0}^t \de s \, M_R(t,s) \, x(s)
 		+ \z_d(t) + \z(t) \ ,
 \\
 & \overline{\z_d (t)} =0 \ , \quad
 \overline{\z_d (t) \z_d(t')}
 	= M_d(t,t') \ ,
 \\
 & \moy{\zeta (t)} =0 \ , \quad
 \moy{\z(t) \z(t')}
 	= \G_C(t,t') + M_C^{(c)} (t,t')
\: .
\end{split}
\label{effective_spin2-bis}
\eeq
Here, the overline represents an average over the Gaussian noise $\z_d(t)$, while the brackets represent an independent average over the Gaussian noise $\z(t)$. 
The first noise represents the fluctuations due to disorder, while the second represents the thermal fluctuations: but at this stage their origin can be forgotten and one can just think to them as independent sources of noise.
Note that this equation is exactly identical to the one of the spherical $p$-spin in zero external field~\cite{BBM96,CC05}, the only difference being that $\z_d(t)=0$ in that case (in presence of an external field $\z_d(t)$ is non zero).

\item
As for the initial condition of Eq.~\eqref{effective_spin2-bis}, we can rewrite Eq.~\eqref{eq-PDF-x0-A-static-cavity}, using Eq.~\eqref{eq-self-consistency-relations1}, as:
\beq
\begin{split}
 & P(x(0)|\z_d(0))
 \propto \exp\left[-\frac{x(0)^2}{2(1-q_g)}+\b_g\z_d(0) x(0)  \right]
 \ ,
 \\
 & P(\z_d(0))
 =\sqrt{\frac{\b_g^2(1-q_g)^2}{2\pi q_g}}\exp\left[-\frac{(1-q_g)^2}{2q_g}\b_g^2\z_d^2(0)\right]
 \ ,
\end{split}
\label{eq-PDF-x0-A-static-cavity-bis}
\eeq
having recognised that the random variable $A$ defined in Eq.~\eqref{A} is nothing but $-\b_g \z_d(0)$ in the large-$N$ limit, according to the definition of this noise in Eq.~\eqref{noises}. Note that a special case is the limit $\b_g=0$ in which $q_g=0$ and one is just extracting the initial condition according to $P(x(0)) = \exp[-x(0)^2/2]/\sqrt{2\pi}$,
\textit{i.e.} uniformly over the sphere.

\item
For the typical reduced gap $r(t)$ the dynamics is given by Eq.~\eqref{eff_gap}:
\beq\label{eq:eff_gap_final}
\begin{split}
 & \int_{0}^t \de s \, \G_R(t,s) \, \dot r(s)
 =  - \tilde \n(t) r(t)
 	- v'(r(t)-w)
 	+\b_g M_{C}^{(c)}(t,0) \, r(0)
 	+ \int_{0}^t \de s \, M_R(t,s)r(s)
 	+ \z_d(t) + \z(t)
 \: ,
\end{split}
\eeq
where $\z(t)$ and $\z_d(t)$ are {\it independent} copies of the same noises introduced in Eq.~\eqref{effective_spin2-bis}, hence with the same
statistical properties.

\item
As for the initial condition of Eq.~\eqref{eq:eff_gap_final}, we can deduce it from Eq.~\eqref{eq-initial-condition-r0-with-q}. If we identify ${\omega/(1-q_g)}$ with ${\beta_g \zeta_d(0)}$ we obtain:
\beq
\begin{split}
 & P(r(0)| \z_d(0))
 \propto \exp\left[-\frac{r(0)^2}{2(1-q_g)} + \beta_g \zeta_d(0) r(0) - \b_gv(r(0)-w)\right]
 \ ,
 \\
 & P(\z_d(0))
 =\sqrt{\frac{\b_g^2(1-q_g)^2}{2\pi q_g}}\exp\left[-\frac{(1-q_g)^2}{2q_g} \b_g^2\z_d^2(0)\right]
 \ .
\end{split}
\label{eq:cavity_prgivenz}
\eeq
Once again in the limit ${\b_g=0}$, ${q_g=0}$, one has ${P(r(0)) = \exp[-r(0)^2/2]/\sqrt{2\pi}}$ provided that
$v(h)$ is a smooth potential.

\item
All these relations depend on the parameter $q_g$, which is self-consistently given by Eq.~\eqref{static_cavity}:
\beq
 \frac{q_g}{(1-q_g)^2}
 	= \a \b_g^2 \int_{-\infty}^\infty \frac{\de \omega \, e^{-\omega^2/(2q_g)}}{\sqrt{2\pi q_g}
 		}
 		\left[ \frac{\int_{-\infty}^\infty \de r \, v'(r-w) \exp\left[-\frac{(r-\omega)^2}{2(1-q_g)} - \b_gv(r-w)\right]}{\int_{-\infty}^\infty \de r \exp\left[-\frac{(r-\omega)^2}{2(1-q_g)} - \b_gv(r-w)\right]}
 		\right]^2
\eeq
so that it depends only on the ratio ${\alpha=M/N}$, the inverse temperature $\beta_g$, and the potential $v(h)$ with its parameter $w$.

\item
The noise and friction kernels are defined self-consistently as averages over the effective process \eqref{eq:eff_gap_final} for ${r(t)}$ or equivalently ${h(t)=r(t)-w}$:
\beq
\begin{split}
 \tilde{\nu}(t)
 & \stackrel{\eqref{eq-cavity1-redef-Lagrange}}{=} 
 	 \hat\nu(t) + \a \overline{\moy{v''(h(t))}}_h
 \ ,
 \\
 M_R(t,t')
 & \stackrel{\eqref{eq-cavity1-III}}{=} 
 	\alpha \, \left. \frac{\delta \overline{ \langle v'(h(t)) \rangle}_h }{\delta P(t')} \right|_{P(t)=0}
 \ ,
 \\
 M_C^{(c)}(t,t')
 & \stackrel{\eqref{eq-cavity1-II-bis}}{=} 
	\alpha \argc{ \overline{\moy{ v'(h(t))v'(h(t'))}}_h - \overline{\moy{v'(h(t))}_h \moy{v'(h(t'))}}_h }
 \ ,
 \\
 M_d(t,t')
 & \stackrel{\eqref{noises-corr-zetad}}{=} 
 	\a \overline{\moy{v'(h(t))}_h \moy{ v'(h(t'))}}_h
 \ ,
\end{split}
 \label{eq:kernels_final}
\eeq
where the brackets are averages over $\z(t)$ and the overlines are averages over $\z_d(t)$, and
the field $P(t)$ has to be added to Eq.~\eqref{eq:eff_gap_final} by replacing ${v'(r(t)-w) \to v'(r(t) + P(t) -w)}$.
Note that for the $p$-spin model, as well as in Mode-Coupling Theory, these kernels can be simply expressed as power-laws of the correlation and response functions, 
which makes the problem much simpler~\cite{Cu02,CC05,RC05}.
Finally, the Lagrange multiplier $\hat \n(t)$ that is contained in $\tilde \n(t)$ must be fixed by the spherical constraint, which implies the condition ${\overline{\langle x^2(t)\rangle}=1}$, that will be solved explicitly in Sec.~\ref{sec-cavity-method-correlation-reponse}.

\end{enumerate}

These findings constitute our main result.
We will see in Sec.~\ref{sec-SUSY} how the same result can be recovered by a different approach based on path integrals.

\subsection{Dynamical equations for the correlation and response functions}
\label{sec-cavity-method-correlation-reponse}

In Sec.~\ref{sec-cavity-method-summary} we have summarised the results of the cavity derivation: a self-consistent equation for the memory kernels $M_R(t,t')$ and ${M_C(t,t') = M_C^{(c)}(t,t') + M_d(t,t')}$, which are written as averages over an effective stochastic process for the typical gap $h(t)$, together with an effective stochastic process, Eq.~\eqref{effective_spin2-bis}, for the typical variable $x(t)$ that depends on these kernels.

From Eq.~\eqref{effective_spin2-bis}, one can derive evolution equations relating the correlation and response functions,
\beq\label{eq:RCdef}
\begin{split}
 \frac1N \sum_{i=1}^N \overline{\moy{x_i(t) x_i(t')}} \stackrel{(N \to \infty)}{\to} \overline{\moy{x(t)x(t')}}
 & \equiv C(t,t') 
 \ , \\
 \frac1N \sum_{i=1}^N \overline{\moy{x_i(t)} \moy{x_i(t')}} \stackrel{(N \to \infty)}{\to} \overline{\moy{x(t)} \moy{x(t')}}
 & \equiv C_d(t,t') 
 \ , \\
 \frac1N \sum_{i=1}^N \frac{\delta \overline{\moy{x_i(t)}}}{\delta H_i(t')} \stackrel{(N \to \infty)}{\to} 
\frac{\delta \overline{\moy{x(t)}}}{\delta H(t')} = 
 \overline{\moy{\frac{\delta x(t)}{\delta \zeta (t')}}}
 & \equiv R(t,t') \ ,
\end{split}
\eeq
to the memory kernels.
Here, for all these observables, the original averages over the microscopic dynamics of Eq.~\eqref{eq:Lang} becomes, in the thermodynamic limit, averages over the effective process of Sec.~\ref{sec-cavity-method-summary}, with the averages over thermal noise and disorder being replaced by the averages over the effective noises.
For the response function, the field $H(t')$ should be added linearly to Eq.~\eqref{effective_spin2-bis}, but because this equation is linear, one can also formally use $\z(t')$ to define the response~\cite{Cu02,CC05}.
Note however that, at ${t \geq t'=0}$, one should not mistake the \emph{dynamical} reponse function ${R(t,0)}$ with the response resulting from a modification of the initial condition itself, which would add an additional contribution that we will not consider thereafter. Note that by causality one has $R(t,t')=0$ for $t'>t$ and in particular $R(t,t' \to t^+)=0$.
Because the derivation of these equation is standard, we only give the main steps and we refer to Refs.~\cite{Cu02,CC05} for more details.

Multiplying Eq.~\eqref{effective_spin2-bis} by ${x(t')}$ with ${t' \in [0,t]}$ and taking the average over the two effective noises ${\zeta_d(t)}$ and ${\zeta(t)}$, we get the following equation for the correlation:
\beq
\begin{split}
 \int_0^t \de s \, \G_R(t,s) \, \partial_s C(s,t')
 =& -\tilde \nu(t)C(t,t') +\b_g M_{C}^{(c)}(t,0)C(t',0)+  \int_{0}^t \de s M_R(t,s) C(s,t')
 \\
 &+\overline{ \z_d(t)\moy{x(t')}} + \overline{\moy{ \z(t)x(t')}}
 \ .
\end{split}
\eeq
Using standard manipulations~\cite{CC05}, or Girsanov theorem, one can show that
\begin{eqnarray}
\overline{\z_d(t) \langle x(t')\rangle}
 &=& \int_0^{t'} \de s \, M_d(t,s) \, R(t',s) 
 		+ \b_g M_d(t,0) \argc{ C(t',0) -C_d(t',0)}
 \ ,
 \\
 \overline{ \langle \z(t) x(t')\rangle }
 &=& \int_0^{t'} \de s \left[ \G_C(t,s) + M_C^{(c)} (t,s)\right]R(t',s)
 \ ,
 \label{eq-average-zetad-x-explicit}
\end{eqnarray}
with the integral boundaries ${s \in [0,t']}$ reflecting the causality encoded in the response ${R(t',s)}$.
%
%
We can note that we have
${M_C^{(c)} (t,s) + M_d(t,s)=\alpha \overline{\moy{ v'(h(t))v'(h(s))}} \equiv M_C(t,s)}$,
\textit{i.e.}~the second moment of the `forces' ${v'(h(t))}$ instead of its second cumulant ${M_C^{(c)}}(t,s)$.
Therefore the dynamical equation for the correlation function becomes
\beq
\begin{split}
 \int_{0}^t \de s \, \G_R(t,s) \, \partial_s C(s,t')
 =&	-\tilde \nu(t)C(t,t')
 	+  \int_{0}^t \de s M_R(t,s) C(s,t')
 	+ \int_0^{t'} \de s \argc{ \G_C(t,s) + M_C^{(c)} (t,s) + M_d(t,s) }R(t',s)
 \\
 &	 	+ \b_g M_{C}^{(c)}(t,0) \, C(t',0)
 		+ \beta_g M_d(t,0) \, \argc{C(t',0)-C_d(t',0)}
 	\ .
\end{split}
\label{Ccavity}
\eeq
In order to obtain the equation for ${C_d(t,t')}$, we first average over $\zeta(t)$, then we multiply by ${\moy{x(t')}}$ with ${t' \in [0,t]}$, and eventually average over $\zeta_d(t)$:
\beq
\begin{split}
 \int_0^t \de s \, \G_R(t,s) \, \partial_s C_d(s,t')
 =&	-\tilde \nu(t)C_d(t,t')
 	+  \int_{0}^t \de s M_R(t,s) C_d(s,t')
 	+ \int_0^{t'} \de s \, M_d (t,s) \, R(t',s)
 \\
 &	 	+ \b_g M_{C}^{(c)}(t,0) \, C_d(t',0)
 		+ \beta_g M_d(t,0) \, \argc{C(t',0)-C_d(t',0)}
 \, .
\end{split}
\label{Ccavity-bis}
\eeq
We see that the structures of both these dynamical equations are very close to each other,
and that they are coupled through the contribution ${\propto \argc{C(t',0)-C_d(t',0)}}$.

With a similar strategy we can write the equation for the response function.
We use the definition of $R(t,t')$ in Eq.~\eqref{eq:RCdef}, from which it follows that
${\partial_s R(s,t')=\overline{\moy{\frac{\delta \dot x(s)}{\delta \z(t')}}}}$.
Differentiating Eq.~\eqref{effective_spin2-bis} with respect to $\z(t')$, we then obtain an equation for $R(t,t')$.
Using that by causality ${R(s,t') \propto \theta(s-t')}$,
and that the initial condition $x(0)$ that appears in Eq.~\eqref{effective_spin2-bis} is independent of $\z(t')$, we get
\beq
\begin{split}
 & \int_{t'}^t \de s \, \G_R(t,s)\partial_s R(s,t')
  = \delta(t-t') -\tilde \nu(t) R(t,t')
	 +\int_{t'}^t \de s M_R(t,s)R(s,t')
 \ .
\end{split}
\label{Rcavity}
\eeq

The Lagrange multiplier can be fixed by the requirement that ${C(t,t) = \overline{\moy{x(t)^2}} = 1}$, as inherited from the spherical constraint.
Evaluating Eq.~\eqref{Ccavity} at ${t=t'}$, one can deduce:
\beq
\begin{split}
 \tilde \n(t)
 = & \int_{0}^t \de s \arga{
 	M_R(t,s) C(s,t) 
 		- \Gamma_R(t,s) \, \partial_s C(s,t)
 		+ \argc{ \G_C(t,s) + M_C (t,s)} R(t,s)}
 \\
 & \quad\quad\quad\quad\quad\quad\quad\quad
 		+\b_g M_C (t,0) C(t,0) - \beta_g M_d(t,0) C_d(t,0)
 \:.
\end{split}
\label{Lag_eq}
\eeq
We have thus obtained the set of coupled Eqs.~\eqref{Ccavity}-\eqref{Ccavity-bis}-\eqref{Rcavity}-\eqref{Lag_eq}, that allow one to obtain the physical observables $C(t,t')$, $C_d(t,t')$ and $R(t,t')$ from the memory kernels, and Eq.~\eqref{Lag_eq} that provides an expression of the Lagrange multiplier that enters in the effective process.
Note that these equations are exactly identical to the ones of the spherical $p$-spin model~\cite{BBM96,CC05},
though the relations between the memory kernels and the response and correlation are very different, as already discussed.

\subsection{The equilibrium case}
\label{sec-cavity-method-equilibrium}

The analysis of the effective stochastic process is highly nontrivial, because the memory kernels are themselves self-consistently determined as averages over this process.
Unfortunately, no analytical treatment is possible and the solution must be found numerically.
Still, the analysis simplifies considerably and can be carried on analytically in the equilibrium case, 
\textit{i.e.}~the steady state with fluctuation-dissipation theorem (FDT) and time-translational invariance (TTI), 
for a thermal bath of inverse temperature ${\beta=1/T}$.
In this case one should recover the results obtained from the static cavity and replica approaches~\cite{BBM96}.
In this section we discuss this connection, which serves as a test of the correctness of the derivation.

Let us consider the equilibrium limit in which the thermal bath that appears in the dynamical equations is stationary and satisfies a FDT in the form ${\G_R(t-t') = \b \th(t-t') \G_C(t-t')}$ (see Sec.~\ref{sec-def-model}), with the same inverse temperature as the one that defines the initial condition ($\b=\b_g$).
In that case the initial Boltzmann measure~\eqref{eq-pdf-initial-condition-equilibrium} is the one corresponding to the equilibrium state of the dynamics, therefore the dynamics must also be stationary: the averages of physical obserables are time independent, while correlation functions $C(t,t')$ become functions of the time-difference $t-t'$, and we thus sometimes write them, with an abuse of notation, as functions of a single time, $C(t,t') \to C(t-t') \to C(t)$.
The correlation-response functions and memory kernels are time-translational invariant, and also satisfy FDT relations:
\beq
 R(t-t')
 	= -\b \th(t-t') \partial_t C(t-t')
 \ , \ \ \ \ \ \ \ \ \ \ \ \ \
 M_R(t-t')
 	= -\b \th(t-t') \partial_t M_C^{(c)}(t-t')
 \ .
\label{eq-FDT-relations}
\eeq
In this case, the dynamics is in equilibrium at all times.
We now restrict to a paramagnetic, or replica symmetric, phase where there is a single pure state.
In this case, at long times, the configuration decorrelates from the initial condition. 
As a consequence, $\la x_i(t) x_i(0) \ra \to \la x_i \ra^2$ for $t\to\io$, and therefore the overlap introduced in Eq.~\eqref{PDF-r0-def-q} can be extracted from the long-time limit of the correlation function, $q_g = q = \lim_{t\to\io} C(t)$.
From this observation, we can recover the results obtained through the static cavity method in Sec.~\ref{sec-cavity-method-initial-condition}.

In the following, to simplify the analysis, we focus on the simplest case ${\Gamma_C(t)=2T\gamma \delta(t)}$, but the results can be easily extended to a generic $\G_C(t)$.
Because the averages of physical obserables are time independent, we have from Eq.~\eqref{eq:kernels_final}:
\beq\begin{split}
 & \tilde \nu(t)
 	= \hat \nu + \alpha \overline{\moy{ v''(h) }}_h
 	\equiv \tilde{\nu}_{\text{eq}}
 \ ,
 \\
 & M^{(c)}_C(t,t) =
 	\alpha [ \overline{ \moy{v'(h)^2}_h - \moy{v'(h)}_h^2} ]
 	\equiv M_C^{(c)}(0)
 \ ,
 \\
 & M_d(t,t') =
 	\alpha \overline{\moy{v'(h)}_h^2}
	\equiv M_d^{\text{eq}}
 \ .
\end{split}\eeq
The last relation implies that ${\zeta_d}$ is a time-independent Gaussian random field with the following statistics:
\beq
 	\overline{ \z_d} = 0
 \ , \quad
 	\overline{\z_d^2} = M_d^{\text{eq}}
 \quad \Rightarrow \quad
 	P(\zeta_d) = \frac{1}{\sqrt{2 \pi M_d^{\text{eq}}}} \exp \arga{- \frac{\zeta_d^2}{2M_d^{\text{eq}}}}
 \ .
\eeq
Furthermore, plugging the FDT relations \eqref{eq-FDT-relations} into Eq.~\eqref{effective_spin2-bis} for the effective variable $x(t)$, into Eq.~\eqref{eq:eff_gap_final} for the effective gap $h(t)$, and into Eq.~\eqref{Ccavity} for the correlation function $C(t)$, one obtains via an integration by parts:
\begin{eqnarray}
 && \gamma \, \partial_t C(t-t') + \beta \int_{t'}^t \de s \, M_C^{(c)}(t-s) \, \partial_{s}C(s-t')
 = 	- \tau_{\text{eq}} \, C(t-t')
 	+ \overline{\z_d \moy{ x } }
 \ ,
 \label{eq-effective-process-equilibrium-1}
 \\
 && \gamma \, \dot x(t) + \beta\int_{0}^t \de s \, M_C^{(c)}(t-s) \, \dot{x}(s)
 =  - \tau_{\text{eq}} \, x(t) + \z_d
 	+ \z(t)
 \ ,
 \label{eq-effective-process-equilibrium-2} 
 \\
  && \gamma\, \dot r(t)+ \beta\int_{0}^t \de s \, M_C^{(c)}(t-s) \, \dot{r}(s)
 =    - \tau_{\text{eq}} \, r(t)
 	- v'(r(t)-w)
 	+ \z_d + \z(t)
 \ ,
 \label{eq-effective-process-equilibrium-2bis} 
 \\
 && \left\lbrace \begin{split}
 	& \overline{\z_d} =0 \ , \quad
 	\overline{\z_d^2}
 	= M_d^{\text{eq}}
 	\ ,
 	\\
 	& \moy{\zeta (t)} =0 \ , \quad
 	\moy{\z(t) \z(t')}
 	= 2 T \gamma \, \delta(t-t') + M_C^{(c)} (t-t')
 	\ ,
 \end{split} \right.
 \label{eq-effective-process-equilibrium-3}
 \\
 && \tau_{\text{eq}} \equiv \tilde \nu_{\text{eq}} - \beta M_{C}^{(c)}(0)
 	= T + \overline{ \z_d \moy{x}}  \ ,
 \label{eq-effective-process-equilibrium-4}
\end{eqnarray}
which are the equations that fully describe the equilibrium dynamics.
The second expression for $\tau_{\text{eq}}$ in Eq.~\eqref{eq-effective-process-equilibrium-4} can be deduced
by taking the limit ${t' \to t^-}$ in Eq.~\eqref{eq-effective-process-equilibrium-1}, using that
${\gamma \, \partial_t C(t-t') \to T}$ for ${t' \to t^-}$.

Let us now analyze these equations. 
The dynamical Eq.~\eqref{eq-effective-process-equilibrium-2} is the one of a harmonic oscillator with
force ${f_x(x) = - \tau_{\text{eq}} \, x + \z_d}$,
which derives from a potential ${V_x(x) = \frac12 \tau_{\text{eq}} \, x^2 + \z_d x}$,
coupled to an equilibrium thermal bath described by the friction term and the noise $\z(t)$.
At all times, $x$ is thus described by its stationary measure proportional to $\exp[-\b V_x(x)]$, \textit{i.e.}
\beq
 P(x|\z_d)
 \propto \exp\left[-\b\left(\frac12 \tau_{\text{eq}} \, x^2 -\z_d \, x \right) \right]
 \ .
\eeq
It follows that $\moy{x} = \int \de x P(x|\z_d) \, x = \zeta_d / \tau_{\text{eq}}$, and
\beq
 \overline{ \z_d \moy{x} }
 = \int \frac{\de \z_d}{\sqrt{2\pi M_d^{\text{eq}} }} \, e^{-\frac{\z_d^2}{2 M_d^{\text{eq}} }}
 \frac{ \z_d^2 }{\tau_{\text{eq}}}
 =\frac{M_d^{\text{eq}}}{\tau_{\text{eq}}}
 \ .
\label{eq-correlation-zetad-x}
\eeq
As a consequence,
\beq
 \tau_{\text{eq}}
 \stackrel{\eqref{eq-effective-process-equilibrium-4}}{=} T + \overline{ \z_d \moy{x}} 
 \stackrel{\eqref{eq-correlation-zetad-x}}{=} T + \frac{M_d^{\text{eq}}}{\tau_{\text{eq}}}
 \ .
 \label{eq:st1}
\eeq
We can obtain a second relation between ${\tau_{\text{eq}}}$ and ${M_d^{\text{eq}}}$ by taking the long-time limit of 
Eq.~\eqref{eq-effective-process-equilibrium-1}.
In this limit, the left hand side vanishes. The first term in the left hand side vanishes because $C(t-t') \to q$ becomes a constant, hence $\partial_t C(t-t') \to 0$.
The second term vanishes because when $t-t'$ is very large, either ${M_C^{(c)}(t-s) \to 0}$ (when $s$ is far from $t$) or ${\partial_{s}C(s-t') \to 0}$ (when $s$ is far from $t'$), or both.
Recalling that ${q = \lim_{t \to \infty}C(t)}$, we thus obtain
\beq
 0=-\tau_{\text{eq}} q+ \overline{\z_d \moy{ x } } = -\tau_{\text{eq}} q +\frac{M_d^{\text{eq}}}{\tau_{\text{eq}}}
 \ .
\label{eq:st2}
\eeq
From Eqs.~\eqref{eq:st1} and \eqref{eq:st2} it follows that
\beq
\begin{split}
 \textit{(i)} \quad
	& \tau_{\text{eq}}
 	=\frac{1}{\beta (1-q)} \ ,
\\
 \textit{(ii)} \quad
 	& \frac{q}{(1-q)^2} = \beta^2 M_d^{\text{eq}} = \a \b^2 \overline{\moy{ v'(h)}^2} 
 \:.
\end{split}
\label{eq-self-consistency-relations-bis}
\eeq
Finally, Eq.~\eqref{eq-effective-process-equilibrium-2bis} describes a particle $r(t)$ moving in a potential $V_r(r) =\frac12 \tau_{\text{eq}} \, r^2 + \z_d r + v(r-w)$,
and in contact with an equilibrium bath, so the stationary measure for $r(t)$ is proportional to $\exp[-\b V_r(r)]$.
Given this observation, and using Eq.~\eqref{eq-self-consistency-relations-bis}, it is easy to check that the distributions $P(r|\z_d)$ and $P(\z_d)$ coincide with the ones obtained in Eq.~\eqref{eq:cavity_prgivenz} in Sec.~\ref{sec-cavity-method-initial-condition}.
As a consequence, the second Eq.~\eqref{eq-self-consistency-relations-bis}, which is a self-consistent equation for $q$, coincides with Eq.~\eqref{static_cavity}, that was obtained through the static cavity method in Sec.~\ref{sec-cavity-method-initial-condition}.
And, without surprise, both equations coincide with what can be obtained from a purely static replica computation \cite{FPSUZ17}.
Moreover, this derivation has provided an alternative physical interpretation to the parameter ${q_g=\overline{\moy{x}^2}}$ as the long-time plateau of the correlation function, since ${q_g=q=\lim_{t \to \infty} C(t)}$.

\section{Derivation of the dynamical mean-field equations through path-integral and supersymmetry}
\label{sec-SUSY}
%

We now follow an alternative and complementary path to obtain the same dynamical equations as via the cavity method, namely the \textit{Martin-Siggia-Rose-Jensen-De Dominicis} (MSRJD) formalism for path integrals in its supersymmetric (SUSY) form.
In the large-$N$ limit these path integrals will be dominated by the saddle point of their SUSY dynamical action, that we will determine.
From there we will show how we can recover the mean-field effective stochastic process summarised in Sec.~\ref{sec-cavity-method-summary}, along with the correlation and reponse dynamical equations given in Sec.~\ref{sec-cavity-method-correlation-reponse}.

Thereafter, we first compute the dynamical MSRJD action of the random continuous perceptron and determine its large-$N$ saddle point exploiting its SUSY formulation (Sec.~\ref{sec-SUSY-action-saddle-pt}).
Secondly we extract from the saddle-point equation the effective stochastic process for the typical reduced gap ${r(t)=h(t)+w}$ (Sec.~\ref{sec-SUSY-stoch-process}).
Thirdly we derive in a similar way the set of dynamical equations for the correlation and response functions, using the SUSY algebra as a powerful shortcut (Sec.~\ref{sec-SUSY-dynamics}).

\subsection{SUSY dynamical action and its large-$N$ saddle point}
\label{sec-SUSY-action-saddle-pt}

The SUSY formulation of the dynamics of Eq.~\eqref{eq:Lang}
can be found in Ref.~\cite{Cu02} but we use slightly different conventions and we do not introduce fermions because 
we use the Ito convention~\cite{kamenev,ZJ02}.
We would like to compute the dynamical action associated to
\beq
 Z_{\text{dyn}}
 \equiv \overline{\moy{
 	\int \DD \vec X(t) \frac{1}{Z(\b_g)} e^{-\b_g \hat{\mathcal{H}} \argp{\vec X(0)}}
 	\prod_{i=1}^N \delta \left[
 	-\int_{0}^t \de s \, \G_R(t,s) \dot x_i(s) -\hat \n(t) x_i(t) - \frac{\partial \mathcal{H} \argp{\vec{X}(t)}}{\partial x_i(t)} +  \h_i(t)
 	\right]
 	}}
 =1
 \label{eq-SUSY-Zdyn-def-1}
\eeq
where the overline and the brackets stand for the average over the quenched disorder and the noise $\vec{\eta}$ respectively, and we have made explicit the average over the initial configuration \cite{HJY83}.
The measure ${\int \DD \vec X(t)}$ thus sums over all the possible histories of the degrees of freedom $\arga{x_i(t)}_{i=1,\dots,N}$ compatible with the Langevin dynamics~\eqref{eq:Lang} and with a stochastic initial condition sampled according to Eq.~\eqref{eq-pdf-initial-condition-equilibrium}.
We have denoted
\beq
 \hat{\mathcal{H}}\argp{\vec X (0)}
 = \frac{\l}{2}\sum_{i=1}^N x_i(0)^2 +\sum_{\m=1}^M v(h_\mu(\vec X (0)))
 \label{eq-SUSY-hat-Hamiltonian-def}
\eeq
where $\l$ is the Lagrange multiplier needed to enforce the spherical constraint specifically on the initial condition as in Eq.~\eqref{eq-pdf-initial-condition-Boltzmann}.

Before going further, it is useful to decompose the kernels $\Gamma_R$ and $\Gamma_C$ into their regular and singular parts:
\beq
 \Gamma_R(t,s) = \gamma_R \delta(t-s) + \theta(t-s) \Gamma_R^*(t,s)
 \ , \qquad
 \Gamma_C(t,s) = 2T\gamma_C \delta(t-s) + \Gamma_C^*(t,s) \ .
\label{eq-GammaRC-singular-regular-parts}
\eeq
This decomposition is useful because we can 
reformulate the friction term that appears in the Dirac $\delta$ function of Eq.~\eqref{eq-SUSY-Zdyn-def-1}, via an integration by parts, as
\beq
 -\int_{0}^t \de s \, \G_R(t,s) \dot x_i(s) -\hat \n(t) x_i(t)
 = \int_{0}^t \de s \, \partial_s \G_R(t,s) \, x_i(s) - \argc{\hat \n(t)+\Gamma_R^*(t,t) } x_i(t) + \Gamma_R^*(t,0) \, x_i(0) \ .
\eeq
One technical pitfall is that for $\Gamma_R$ both its singular and regular parts have a discontinuity at ${s=t}$, which 
has to be considered with care when performing integrations by parts.
In fact the integration boundaries are always to be understood as ${s \in [0,t^+]}$, so that there is no ambuguity on cutting half of Dirac $\delta$ or Heaviside $\theta$ functions.
Note that we can recover the case of a white equilibrium thermal bath by simply imposing ${\gamma_R=\gamma_C=\gamma}$, ${\Gamma_R^* \equiv 0}$ and ${\Gamma_C^* \equiv 0}$ at any step in the computation, and we can also recover the case without a white noise by setting $\g_R = \g_C =0$.
The alternative would be to consider
${\Gamma_R(t)=\gamma \frac{1}{\tau} e^{-t/\tau} \theta(t)}$
and ${\Gamma_C(t) = T \gamma \frac{1}{\tau} e^{-t/\tau} }$
in the limit ${\tau \to 0}$, but this is less straightforward especially for intermediary steps.

\subsubsection{Introducing replicas for the initial condition}
\label{sec-SUSY-action-saddle-pt-replicas-initial-condition}

Because the partition function $Z(\b_g)$ depends on the quenched disorder, to treat the initial condition we need to introduce replicas, \textit{i.e.}~consider now ${\lbrace \vec{X}^{(\sigma)}(t) \rbrace}$ with ${\sigma=1, \dots, n}$ along with their associated response fields ${\lbrace \hat{\vec{X}}^{(\sigma)}(t) \rbrace}$:
\beq
\begin{split}
 Z_{\mathrm{dyn}}
 &= \lim_{n\to 0} \overline{\moy{
 	\int \DD \vec X^{(1)}(t) \, {Z(\b_g)}^{n-1}e^{-\b_g \hat{\mathcal{H}} \argp{\vec X ^{(1)} (0)}}
 	\prod_{i=1}^N \delta \argp{
 		-\int_{0}^t  \!\! \de s \, \G_R(t,s) \, \dot{x}_i ^{(1)} (s) - \hat \n(t) \, x_i ^{(1)}(t) - \frac{\delta \mathcal{H} \argp{\vec{X} ^{(1)}(t)}}{\delta x_i ^{(1)}(t)} + \h_i(t)
 	}}}
 \\
 &=\lim_{n\to 0} \int \argp{ \prod_{\sigma=2}^n \de \vec X^{(\sigma)}(0)} \int \DD \vec X^{(1)}(t) \DD \hat{\vec{X}}^{(1)}(t)\, e^{ S_{\mathrm{dyn}}}
\end{split}
\label{eq-Zdyn-very-first-version}
\eeq
where the dynamical action $S_{\mathrm{dyn}}$ is obtained by standard manipulation using that the ${\eta_i (t)}$ are Gaussian noises in Eq.~\eqref{eq:Lang}:
\beq
\begin{split}
 S_{\mathrm{dyn}}
 =& \int_0^{\infty} \de t \int_0^{\infty} \de t' \left[
 	\frac12   \G_C(t,t') \, i\hat {\vec X}^{(1)}(t)  \cdot  i\hat{ \vec X}^{(1)}(t')
 	+  \partial_{t'} \G_R(t,t') \, i \hat{ \vec X}^{(1)}(t) \cdot {\vec X}^{(1)}(t') \right]
 \\
 & - \int_0^{\infty} \de t \, \argc{\hat \n(t)+ \Gamma_R^*(t,t)} \, i\hat{ \vec X}^{(1)}(t) \cdot \vec X^{(1)}(t)
	+ \int_0^{\infty} \de t  \, \Gamma_R^*(t,0) \, i\hat{\vec{X}}^{(1)}(t) \cdot \vec{X}^{(1)}(0)
 \\
 & + \ln \overline{\exp\left[
 	-\int_0^{\infty} \de t  \, i\vec{\hat X}^{(1)}(t) \cdot \frac{\partial \mathcal{H} \argp{\vec{X}^{(1)}(t)}}{\partial \vec X^{(1)}(t)}
 	- \b_g \sum_{\sigma=1}^n \hat{\mathcal{H}} \argp{\vec{X}^{(\sigma)}(0)}
 \right]}
\end{split}
\eeq

At this point we can note that we can make all the replicas become dynamical.
This is due to the fact that for any realisation of the initial condition and disorder we have
\beq
 1 =
 \int \DD \vec X(t) \, \delta \argp{
 	\int_{0}^t \de s \, \partial_s \G_R(t,s) \, {\vec X}(s)
 	- \argc{ \hat \n(t) + \Gamma_R^*(t,t) } \vec X(t)
 	+ \Gamma_R^*(t,0) \, \vec X(0)
 	- \frac{\delta \mathcal{H} \argp{\vec{X}(t)}}{\delta \vec X(t)}
 	+ \vec \eta(t)
 }
 \:.
\eeq
Therefore we can write
\beq
 Z_{\mathrm{dyn}}
 = \lim_{n \to 0} \int \argp{ \prod_{\s=1}^n \DD \vec{X}^{(\s)}(t) \DD \hat{\vec X}^{(\s)}(t) }
 	\, e^{ S_{\mathrm{dyn}}}
=1 \ ,
\label{eq-dynamical-action-expression-1}
\eeq
where we have redefined
\beq
\begin{split}
 S_{\mathrm{dyn}}
 =& \sum_{\s=1}^n 
 	\int_0^{\infty} \de t \, \de t' \left[
 		\frac12 \G_C(t,t') \, i\hat {\vec X}^{(\s)}(t) \cdot i\hat{ \vec X}^{(\s)}(t')
 		+ \partial_{t'} \G_R(t,t') \, i \hat{ \vec X}^{(\s)}(t) \cdot {\vec X}^{(\s)}(t') \right]
 \\
 & - \sum_{\s=1}^n \int_0^{\infty} \de t \, \argc{\hat \n(t)+\Gamma_R^{*}(t,t)} \, i\hat{ \vec X}^{(\s)}(t) \cdot \vec X^{(\s)}(t)
  	+ \sum_{\s=1}^n \int_0^{\infty} \de t \, \Gamma_R^{*}(t,0) \, i\hat{ \vec X}^{(\s)}(t) \cdot \vec X^{(\s)}(0)
 \\
 & - \sum_{\s=1}^n \frac12 \beta_g \lambda \cdot \vec{X}^{(\sigma)}(0)^2 
 + \log \overline{\exp \arga{ - \sum_{\mu=1}^{M} \sum_{\s=1}^n \argc{ 
	 	\int_0^{\infty} \de t \, v' \argp{ h_\mu^{(\sigma)}(t)} \, i\vec{\hat X}^{(\s)}(t) \cdot \vec{F}^{\mu}
 		+ \b_g  v \argp{ h_\mu^{(\sigma)}(0)}
 	 }}}
 	\ .
\end{split}
\label{eq-Sdyn-before-SUSY}
\eeq
We used Eq.~\eqref{eq-SUSY-hat-Hamiltonian-def} to make explicit $\hat{\mathcal{H}}(\vec{X}^{(\sigma)}(0))$.
We recall that the disorder corresponds to the random Gaussian constraints ${\lbrace \vec{F}_{\mu} \rbrace_{\mu=1,\dots,M}}$ defining the gaps ${h_{\mu}^{(\sigma)}(t) = \vec{F}_{\mu} \cdot \vec{X}^{(\sigma)}(t) - w}$, hidden in the Hamiltonian potential $v$.
%

\subsubsection{SUSY formulation}
\label{sec-SUSY-action-saddle-pt-SUSY}

The dynamical action can be put in a supersymmetric form, having then a structure similar to the replicated Hamiltonian in the static case \cite{Ku92, ZJ02} and much more compact,
which greatly simplifies the derivation of the saddle point.
In order to do so, we introduce superfields with Grassmann variables $\ththbar{}$ \cite{ZJ02}
\beq
\begin{split}
  &\vec X^{(\s)}(a)
  	= \vec X^{(\s)}(t_a)  + \ththbar{a} \, i \hat {\vec X}^{(\s)}(t_a)
 \ , 
 \\
 & \d(a,b)
 	=  \d(t_a - t_b) (\ththbar{a} + \ththbar{b})
 \ , \ \ \ \ \ \ \ \ \ 
 \d_{\s\t}(a,b)
 	= \d_{\s\t} \d(a,b)
 \ ,
 \\
 & \Rightarrow \int da \, f[ \vec X^{(\s)}(a)]
 	= \int_0^{\infty} \de t_a \, \de \bth_a \de \th_a \, f[ \vec X^{(\s)}(t_a)  + \ththbar{a} \, i \hat {\vec X}^{(\s)}(t_a)]
	= \int_0^{\infty} \de t_a \, i\hat {\vec X}^{(\s)}(t_a) f'[\vec X^{(\s)}(t_a)]
\end{split}
\eeq
and specifically for the SUSY representation of our dynamics we define:
\beq
\begin{split}
 & \b(a) = 1 + \ththbar{a} \, \b_g \, \delta(t_a)
 \ ,
 \\
 & \nu(a) =\hat \nu(t_a) + \Gamma_R^*(t_a,t_a) + \ththbar{a} \, \beta_g \lambda \, \delta(t_a)  
 \ , 
 \\
 &\G_{\s \t}(a,b)
 	= \delta_{\s\t} \underbrace{\left[
	 	\G_C(t_a,t_b) + \ththbar{a} \, \partial_{t_a} \G_R(t_b,t_a) + \ththbar{b} \, \partial_{t_b} \G_R(t_a , t_b)
	\right]}_{\equiv \hat \G(a,b) }
		+ \delta_{\s\t} \argc{\ththbar{a} \delta(t_a) \Gamma_R^*(t_b,t_a) + \ththbar{b} \delta(t_b) \Gamma_R^*(t_a,t_b)}
 \ ,
 \\
 & \KK_{\s\t}(a,b) =- \G_{\s\t}(a,b) + \nu(a) \d_{\s\t}(a,b)
 \:.
\label{eq-def-SUSY-representation-list}
\end{split}
\eeq
We choose as an overall convention that the argument indicates if we are considering a superfield or a scalar, for instance $\vec{X}(a)$ or ${\vec{X}(t_a)}$.
We emphasise that the initial condition is implemented here through the SUSY inverse temperature ${\beta(a)}$, the Lagrange multiplier $\nu(a)$ and the superkernel ${\Gamma_{\sigma \tau}(a,b)}$.
Moreover the latter contains both the friction and noise kernels of the model,
so that combined with the Lagrange multiplier into the kinetic superkernel ${\KK_{\s\t}(a,b)}$ it allows to rewrite the first line of Eq.~\eqref{eq-Sdyn-before-SUSY} in a quadratic form, diagonal in the replica indices.
Indeed, we have that
\beq
\begin{split}
 & \frac12 \sum_{\s,\t=1}^{n} \int \de a \de b \, \G_{\s \t}(a,b) \vec X^{(\s)}(a)  \cdot  \vec X^{(\t)}(b) 		
 \\
 = & \sum_{\s=1}^n \int_0^{\infty} \!\!\!\! \de t \int_0^{\infty} \!\!\!\! \de t' \left[
		\frac12 \G_C(t,t') i\hat {\vec X}^{(\s)}(t) \cdot i\hat{ \vec X}^{(\s)}(t')
		+ \partial_{t'} \G_R(t,t') i \hat{ \vec X}^{(\s)}(t) \cdot {\vec X}^{(\s)}(t')
	\right]
 \\
 	& + \sum_{\s=1}^n \int_0^{\infty} \!\!\!\! \de t \, \G_R^*(t,0) \,  i \hat{ \vec X}^{(\s)}(t) \cdot {\vec X}^{(\s)}(0) \ ,
\end{split}
\eeq
and
\beq\begin{split}
 \frac12 \sum_{\s,\t=1}^{n} \int \de a \de b \, \nu(a) \d_{\s\t}(a,b)
 		\, \vec X^{(\s)}(a)  \cdot  \vec X^{(\t)}(b)
 = & \sum_{\sigma =1}^{n} \int_0^{\infty} \!\!\!\! \de t_a \, \argc{\hat{\nu}(t_a)+\Gamma_R^*(t_a,t_a)} \,
 		i \hat{\vec{X}}^{(\sigma)}(t_a) \cdot \vec{X}^{(\sigma)}(t_a)
 \\
	& - \sum_{\sigma =1}^{n} \frac12 \beta_g \lambda \, \vec{X}^{(\sigma)}(0)^2 \ .
\end{split}\eeq
Combining these last two contributions and defining moreover
\beq
 Q_{\s\t}(a,b)
 = \frac{1}{N} \vec X^{(\s)}(a)\cdot \vec X^{(\t)}(b)
\label{eq-def-Q-sigmatau-ab}
\eeq
and ${\vec{Q}=\arga{Q_{\sigma \tau}}_{\sigma,\tau =1 \dots n}}$,
we can finally rewrite the dynamical action in Eq.~\eqref{eq-dynamical-action-expression-1} as an extensive function in $N$:
\beq
\begin{split}
 & \int \argp{ \prod_{\s=1}^n \DD \vec{X}^{(\s)}(t) \DD \hat{\vec X}^{(\s)}(t) }
 	\, e^{ S_{\mathrm{dyn}}}
 = \int \DD \vec{Q}	 \, e^{ N S(\vec{Q})}
 \\
 & S(\vec{Q})
 =- \frac12 \sum_{\s,\t =1}^n \int \de a \de b \, \KK_{\s\t}(a,b) Q_{\s\t}(a,b) + \frac12 \log\det \vec{Q} + \a \log \ZZ(\vec{Q})
\end{split}
\label{eq-dynamical-action-expression-2}
\eeq
where the $\log\det \vec{Q}$ term comes from the change of variable from $\vec{X}^{(\s)}(a)$ to $\vec{Q}_{\s\t}(a,b)$~\cite{FPSUZ17}, and
\beq
 \alpha \log \ZZ(\vec{Q})
	\stackrel{\eqref{eq-Sdyn-before-SUSY}}{\equiv}
	\frac{1}{N} \log \overline{\exp \arga{ -\sum_{\mu=1}^{M} \sum_{\s=1}^n \argc{ 
	 	\int \de a \, \beta (a) \, v\argp{\vec{X}^{(\sigma)}(a) \cdot \vec{F}^\mu -w}}
 	 }} \ .
\label{eq-def-Zeta-Q-before}
\eeq
The disorder average over the Gaussian random constraints can be taken explicitly, following the same steps as in the static case~\cite{FPSUZ17},
so that we obtain
\beq
  \ZZ(\vec{Q}) = \int \DD_\vec{Q} \vec{r}  \, \Psi(\vec{r})
  \ , \quad \text{with} \quad
 \left\lbrace \begin{array}{l}
 \DD_{\vec{Q}}\vec{r} \propto \DD \vec{r} \, \exp \arga{-\frac12\sum_{\s,\t =1}^n \int \de a \de b \, r_\s(a)  Q^{-1}_{\s\t}(a,b) r_\t(b)}
 \\ \\
 \Psi(\vec{r}) = \exp \arga{ -\sum_{\s=1}^n \int \de a \b(a) \, v \argp{ r_\s(a) - w} }
 \end{array} \right.
\label{eq-def-Zeta-Q}
\eeq
the measures ${\DD_Q \vec{r}}$ and ${\DD \vec{r}}$ summing over all the histories of the $n$ replicas ${\vec{r}=\lbrace r_1, \dots,r_n \rbrace}$ and the measure $\DD_Q \vec{r}$ being normalised to ${\mathcal{N}_Q \equiv \int \DD_Q \vec{r}}$.
This is the first occurrence in this SUSY derivation of what looks like the effective stochastic process we are aiming at, expressed here in terms of the SUSY and replicated 
reduced gap ${r_\s(a)=r_\s(t_a) + \ththbar{a} i \hat{r}_\s(t_a)}$.
Beware that from now on, $r$ will denote this SUSY variable ${r(a)}$, except if specified otherwise through its argument ${r(t_a)}$.
For the time-being it has been introduced as a mathematical trick in order to treat the disorder average in Eq.~\eqref{eq-def-Zeta-Q-before}, but we can already define its corresponding statistical average:
\beq
\moy{\OO(r_\s)}_{\vec{r}}
 = \frac{\int \DD_\vec{Q} \vec{r} \, \OO(r_\s) \, \Psi(\vec{r}) }{\int \DD_\vec{Q} \vec{r} \, \Psi(\vec{r})}
 \ .
\label{eq-def-average-SUSY-r}
\eeq
We emphasise moreover that in the dynamical action, the terms that are \emph{not} diagonal in the replica indices ${(\sigma,\tau)}$ are 
on the one hand ${\frac12 \log \det \vec{Q}}$ and on the other hand the effective dynamical partition function $\ZZ(\vec{Q})$ associated to the 
measure ${\mathcal{D}_\vec{Q} \vec{r}}$.

\subsubsection{Saddle-point equation}
\label{sec-SUSY-action-saddle-pt-equation}

We have thus reformulated the dynamical action as ${S_{\text{dyn}}(\vec{Q})=N S(\vec{Q})}$ with Eqs.~\eqref{eq-dynamical-action-expression-1}, \eqref{eq-dynamical-action-expression-2} and~\eqref{eq-def-Zeta-Q},
making use of the SUSY representation.
Because $S_{\text{dyn}}(\vec{Q})$ is given by $N$ times a function of a finite number of degrees of freedom, 
the large-$N$ limit we are interested in is controlled by the saddle-point value of ${S(\vec{Q})}$:
\beq
 1 = Z_{\mathrm{dyn}}
 =\int \mathcal{D} \vec{Q} \, e^{N S(\vec{Q})}
 \stackrel{(N \to \infty)}{\sim} e^{N S(\widetilde{\vec{Q}})}
 \quad \text{with} \quad
 \frac{\delta S(\vec{Q})}{\delta Q_{\sigma \tau}(a,b)} \Big\vert_{\vec{Q}=\vec{\widetilde{Q}}} =0 \ .
\eeq
We emphasise the physical interpretation of this saddle-point solution, as the mean-field prediction given by the thermodynamic limit of Eq.~\eqref{eq-def-Q-sigmatau-ab}:
\beq
 \widetilde{Q}_{\sigma \tau}(a,b)
 = \lim_{N \to \infty}  \overline{\moy{ \frac{1}{N} \vec{X}^{(\sigma)}(a) \cdot \vec{X}^{(\tau)}(b) }}
 \ .
\label{eq-def-saddle-point-mean-field}
\eeq
The saddle-point equation that gives $\widetilde{Q}_{\sigma \tau}(a,b)$ is directly obtained from Eq.~\eqref{eq-dynamical-action-expression-2} with functional derivatives:
\beq
 0 = - \KK_{\s\t}(a,b) + Q^{-1}_{\s\t}(a,b)
 + \frac{2 \a}{\ZZ(\vec{Q})} \frac{\delta \ZZ (\vec{Q})}{\delta Q_{\s\t}(a,b)} \ .
\eeq
Moreover from Eqs.~\eqref{eq-def-Zeta-Q}-\eqref{eq-def-average-SUSY-r} we deduce~\cite{MKZ16b}
\beq
\begin{split}
 \frac{2 \alpha}{\ZZ(\vec{Q})}\frac{\delta \ZZ (\vec{Q})}{\delta Q_{\s\t}(a,b)}
 &= \b(a) \b(b) \alpha \moy{ v'(r_{\s}(a) - w) v'(r_{\t}(b) - w)}_r
 	- \b(a) \alpha \moy{ v''(r_{\s}(a) - w)}_r \d_{\s\t}(a,b)
 \\
 &\equiv \beta(a) \beta(b) \, M_{\s\t}(a,b) - \beta(a) \, \d\n_\s(a) \, \d_{\s\t}(a,b)
 \ ,
\end{split}
\eeq
with the following definitions
\beq
 \begin{split}
 M_{\s\t}(a,b)
 &= \a \moy{v'(r_{\s}(a) - w) v'(r_{\t}(b) - w)}_r
 \ ,
 \\
 \d\n_\s(a)
 &= \a \moy{v''(r_{\s}(a) -w)}_r
 \ ,
\end{split}
\label{eq:MSUSYdef}
\eeq
where ${M_{\s\t}(a,b) }$ is the memory superkernel and $\d\n_\s(a)$ a correction to the Lagrange multiplier.
Therefore we obtain that the saddle-point ${\widetilde{Q}_{\sigma \tau}(a,b)}$ must satisfy the equation:
\beq
 \widetilde{Q}_{\sigma \tau}^{-1}(a,b)
	= \KK_{\s\t}(a,b)+ \beta(a) \, \d\n_\s(a) \, \d_{\s\t}(a,b) - \beta(a)\beta(b) \, M_{\s\t}(a,b)
	\equiv \LL_{\s\t}(a,b) - \beta(a)\beta(b) \, M_{\s\t}(a,b) \ ,
\label{eq:HSSUSY}
\eeq
where we have separated from the memory superkernel the contribution:
\beq
 \LL_{\s\t}(a,b)
 \equiv  \KK_{\s\t}(a,b) +  \beta(a) \, \d\n_\s(a) \, \d_{\s\t}(a,b)
 \stackrel{\eqref{eq-def-SUSY-representation-list}}{=}
 	-\G_{\s\t}(a,b) + [\nu(a)+ \beta(a) \, \d\n_\s(a)] \, \d_{\s\t}(a,b)
\ ,
\label{eq:HSSUSY-operator-L}
\eeq
diagonal in the replica indices, because it includes explicitly $\delta_{\sigma \tau}$.

The saddle-point solution given by Eqs.~\eqref{eq:HSSUSY}-\eqref{eq:HSSUSY-operator-L} has two complementary consequences.
The first one is the following closure relation:
\beq
\begin{split}
 \delta_{\sigma \tau}(a,b)
 &= \sum_{\gamma=1}^n \int \de c \, \widetilde{Q}^{-1}_{\sigma \gamma}(a,c) \widetilde{Q}_{\gamma \tau}^{\phantom{1} \!}(c,b)
 \\
 &= - \sum_{\gamma=1}^n \int \de c \, \argc{ \Gamma_{\sigma \gamma}(a,c) + \beta(a)\beta(c) M_{\sigma \gamma}(a,c)} \widetilde{Q}_{\gamma \tau}^{\phantom{1} \!}(c,b)
 	+ \argc{\n(a)+ \beta(a) \d\nu_\s(a) } \widetilde{Q}_{\sigma \tau}^{\phantom{1} \!}(a,b) \ .
\end{split}
\label{eq:HSSUSY3}
\eeq
The second one is that the average $\moy{\OO (r_\s)}_\vec{r}$ is given in the large-$N$ limit by substituting ${\vec{Q}=\vec{\widetilde{Q}}}$,
given by Eq.~\eqref{eq:HSSUSY},
 into the measure ${\mathcal{D}_{\vec{Q}} \vec{r}}$ of Eq.~\eqref{eq-def-average-SUSY-r}:
\beq
\begin{split}
 & \lim_{N \to \infty} \moy{ \OO(r_\s) }_\vec{r}
  =	\frac{\int \mathcal{D} \vec{r} \, \OO(r_\s) \, e^{S_{\text{eff}}}}
	{\int \mathcal{D} \vec{r} \, e^{S_{\text{eff}}}}
 \ , \quad
 \mathcal{N}_{\widetilde{Q}} \equiv \int \mathcal{D}_{\widetilde{Q}} \vec{r} =1
 \ ,
 \\
 & S_{\text{eff}}(r_1, \dots, r_n)
 	\equiv 	-\frac12 \sum_{\s,\t=1}^n \int \de a \de b \, r_\s(a) \argc{ \LL_{\s\t}(a,b) - \beta(a)\beta(b) M_{\s\t}(a,b)} r_\t(b)
		-\sum_{\s=1}^n \int \de a \, \b(a) \, v \argp{r_\s(a) -w} \ .
\end{split}
\label{eq:avSUSY}
\eeq
We immediately see that the nontrivial part of this saddle-point solution will be the self-consistent determination of the memory superkernel~\eqref{eq:MSUSYdef}, in the same spirit as pointed out in Sec.~\ref{sec-cavity-method}.

The two equations~\eqref{eq:HSSUSY3}-\eqref{eq:avSUSY} are the large-$N$ dynamical equations in SUSY form,
along with the different definitions 
for ${\lbrace
\beta(a), \nu(a), \Gamma_{\sigma \tau} (a,b), \mathcal{K}_{\sigma \tau}(a,b),
Q_{\sigma \tau}(a,b),
M_{\sigma \tau}(a,b),
\delta \nu_\sigma (a),
\widetilde{Q}_{\sigma \tau}(a,b),
\mathcal{L}_{\sigma \tau}(a,b)
\rbrace}$
given respectively in Eqs.~\eqref{eq-def-SUSY-representation-list}, \eqref{eq-def-Q-sigmatau-ab}, \eqref{eq:MSUSYdef}, \eqref{eq:HSSUSY} and~\eqref{eq:HSSUSY-operator-L}.
Note that only ${M_{\sigma \tau}(a,b)}$ and ${\widetilde{Q}_{\sigma \tau}(a,b)}$ have non-zero contributions for different replicas ${\sigma \neq \tau}$.
In Sec.~\ref{sec-SUSY-dynamics} we will derive the dynamical equations for the correlation and response functions from Eq.~\eqref{eq:HSSUSY3}.
But before, in the next section we will deduce the effective stochastic process corresponding to Eq.~\eqref{eq:avSUSY}.

\subsection{Self-consistent effective stochastic process}
\label{sec-SUSY-stoch-process}

The starting point of the SUSY path-integral formulation is the Langevin dynamics and its initial condition encoded in ${Z_{\text{dyn}}=1}$ in Eq.~\eqref{eq-SUSY-Zdyn-def-1}.
Now that we have successfully performed the averages over noise and disorder to obtain the effective action $S_{\text{eff}}$ in Eq.~\eqref{eq:avSUSY}, 
we want to go the other way around and determine its corresponding effective dynamics on $r(t)$.
Our aim is to show that the average over $\vec{r}$ with the effective action $S_{\text{eff}}$ can be represented, as for the original average, 
into an average over some effective stochastic process with a thermal noise independent for each replica,
denoted by $\moy{\cdots}$, followed by an average over a disorder, denoted by $\overline{\cdots}$, that couples different replicas.

To this aim, we will first make some general remarks on the structure of the replica-SUSY correlator, and then use this structure to simplify the effective action
$S_{\text{eff}}$. Finally we will show that $S_{\text{eff}}$ can be represented by an effective stochastic process as stated above.

\subsubsection{Structure of the replica-SUSY correlators}

We start with some preliminary remarks.
For a given replica $\sigma$ and an analytic function $f(r)$, we have in the MSRJD formalism~\cite{kamenev} that:
\beq\label{eq:pepepe}
\begin{split}
 \moy{f(r_\sigma(t)) \, i\hat{r}_\sigma(t)}_{\vec{r}}=0
 \, , \quad
  \moy{f(r_\sigma(t))f(r_\sigma(t'))  \, i\hat{r}_\sigma(t)i\hat{r}_\sigma(t')}_{\vec{r}}=0
 \ .
\end{split}
\eeq
As for a pair of distinct replicas ${\sigma \neq \tau}$, we have in the initial dynamics Eq.~\eqref{eq:Lang} that the replicas are uncorrelated at fixed disorder, so that we have
${\overline{\moy{ f(\vec{X}^{(\sigma)}(t)) f(\vec{X}^{(\tau)}(t'))}}
=\overline{\moy{ f(\vec{X}^{(\sigma)}(t))} \moy{ f(\vec{X}^{(\tau)}(t'))} }}$. As a consequence of this and Eq.~\eqref{eq:pepepe}, an average of the form
\beq
\overline{\moy{ f(\vec{X}^{(\sigma)}(a)) f(\vec{X}^{(\tau)}(b))}} = \overline{\moy{ f(\vec{X}^{(\sigma)}(a))} \moy{ f(\vec{X}^{(\tau)}(b))}} = 
\overline{\moy{ f(\vec{X}^{(\sigma)}(t_a))} \moy{ f(\vec{X}^{(\tau)}(t_b))}}
\eeq
with $\s\neq \t$
has no terms associated to the Grassman variables $\th_a$ and $\th_b$. In particular, this implies, according to Eq.~\eqref{eq-def-saddle-point-mean-field}, that
for $\s\neq \t$, $\wt Q_{\s\t}(a,b) = \wt Q_{\s\t}(t_a,t_b)$, and using Eq.~\eqref{eq:HSSUSY3} one can check that the same is true for the memory function,
\textit{i.e.} $M_{\s\t}(a,b) = M_{\s\t}(t_a,t_b)$
for $\s\neq \t$.
The diagonal term instead has terms associated to the Grassman variables, but due to Eq.~\eqref{eq:pepepe}, the term with four Grassman variables
$\ththbar{a} \ththbar{b}$ vanishes.

Using these properties, and recalling that we will ultimately replace the average $\moy{ \cdots }_{\vec{r}}$ by a double average $\overline{\moy{\cdots}}$,
where $\moy{\cdots}$ is done independently for each replica,
we can make explicit first the memory superkernel
\beq
\begin{split}
 & M_{\s\t}(a,b)
 \stackrel{\eqref{eq:MSUSYdef}}{=}
 	\argc{ M_C(t_a,t_b) + \ththbar{a} \, M_R(t_b,t_a) + \ththbar{b} \, M_R(t_a,t_b) }
\, \delta_{\sigma \tau} + M_d (t_a,t_b) \, (1-\delta_{\sigma \tau})
 	 \\
 & \phantom{M_{\s\t}(a,b)} \; = \;
 	\argc{ M_C^{(c)}(t_a,t_b) + \ththbar{a} \, M_R(t_b,t_a) + \ththbar{b} \, M_R(t_a,t_b)} \, \delta_{\sigma \tau} + M_d (t_a,t_b)
 	\\
 &	\phantom{M_{\s\t}(a,b)} \; \equiv \; \MM(a,b) \, \delta_{\sigma \tau} + M_d (t_a,t_b)  \ ,
 \\
 & \text{with} \quad
 \left\lbrace \begin{array}{ccl}
 	M_C(t,t')
 		& \equiv & \alpha \overline{\moy{v'\argp{r_\sigma(t)-w} v'\argp{r_\sigma(t')-w}}}
		 			= M_C(t',t) \ ,
 	\\
 	M_d(t,t')
 		& \equiv & \alpha \overline{ \moy{v'\argp{r_\sigma(t)-w}}_r \moy{v'\argp{r_\tau(t')-w}} }
 			= M_d(t',t) \ ,
 	\\
 	M_R (t,t')
 		& \equiv & \alpha \overline{ \moy{v'\argp{r_\sigma(t)-w} v''\argp{r_\sigma(t')-w} i \hat{r}_\sigma(t')}} \ ,
 	\\
 	M_C^{(c)}(t,t') & \equiv & M_C(t,t') - M_d(t,t') \ ,
 \end{array} \right. 
\end{split}
\label{eq-memory-superkernel-explicit}
\eeq
and secondly the correction to the Lagrange multiplier
\beq
\begin{split}
 \d\n_\s(a)
 \stackrel{\eqref{eq:MSUSYdef}}{=} \a \moy{v''(r_{\s}(a) -w)}_{\vec{r}}
 = \a \overline{\moy{v''(r_{\s}(t_a) -w)}}
 =\d\n (t_a)\ ,
\end{split}
\label{eq-correction-SUSY-Lagrange-explicit}
\eeq
where we have also dropped the replica index using the replica symmetry, that is a consequence of our choice of replica symmetric initial condition.
Note that Eqs.~\eqref{eq-memory-superkernel-explicit}-\eqref{eq-correction-SUSY-Lagrange-explicit}
are strongly reminiscent (on purpose) of the definitions~\eqref{eq:kernels_final} given in Sec.~\ref{sec-cavity-method-summary}.
We emphasise moreover that by causality, we have in fact ${M_R(t,t') \propto \theta (t-t')}$ in the MSRJD formalism.

\subsubsection{Structure of the effective action}

We now recall that the initial condition is encoded in our SUSY representation into $\beta(a)$, ${\nu(a)}$ and ${\Gamma_{\sigma \tau}(a,b)}$ given in Eq.~\eqref{eq-def-SUSY-representation-list}.
Defining
\beq
 \nu_0(t)=\hat{\nu}(t) + \Gamma_R^*(t,t)+\delta\nu(t)
 \quad \Rightarrow \quad
 \nu(a)+\delta \nu(a) = \nu_0(t_a) + \ththbar{a} \, \beta_g\lambda \, \delta(t_a) \ ,
\label{eq-def-nu-0-scalar-SUSY}
\eeq
we can now write explicitly the three contributions to the effective action $S_{\text{eff}}$:
\textit{(i)}~a pure dynamical part oblivious of the initial condition,
\textit{(ii)}~a dynamical part which remembers the initial condition,
and \textit{(iii)}~the part encoding the distribution of the stochastic initial condition.
Indeed, we obtain with some additional manipulations that
${S_{\text{eff}}=S_{\text{eff}}^{(t)} + S_{\text{eff}}^{(t,0)} + S_{\text{eff}}^{(0)}}$
with
\beq
\begin{split}
 S_{\text{eff}}^{(t)}
 = \frac12 \sum_{\sigma,\tau =1}^n \int \de a \de b \, r_\sigma(a) r_\tau(b) \,
 		&\arga{\delta_{\sigma \tau} \argc{\hat\Gamma (a,b) - \nu_0(t_a) \delta(a,b)} + M_{\sigma \tau}(a,b) }
 	- \sum_{\sigma=1}^n \int \de a \, v\argp{r_\sigma(a)-w} \ ,
 \\
 S_{\text{eff}}^{(t,0)}
 = \frac12 \sum_{\sigma,\tau =1}^n \int \de a \de b \, r_\sigma(a) r_\tau(b) \,
 	& \left\lbrace \delta_{\sigma \tau} \argc{\ththbar{a} \, \delta(t_a) \, \Gamma_R^*(t_b,t_a) + \ththbar{b} \, \delta (t_b) \, \Gamma_R^*(t_a,t_b)} \right.
 \\
 	& \quad \left. + \argc{\ththbar{a} \, \delta(t_a) + \ththbar{b} \, \delta(t_b)} \, \beta_g M_{\sigma\tau}(a,b) \right\rbrace \ ,
 \\
 S_{\text{eff}}^{(0)}
 = \frac12 \sum_{\sigma,\tau =1}^n \int \de a \de b \, r_\sigma(a) r_\tau(b) \,
 	& \ththbar{a} \, \ththbar{b} \, \argc{- \delta_{\sigma\tau} \, \beta_g \argp{\lambda + \delta \nu (0)} \delta (t_a-t_b) \, \delta (t_a) + \beta_g^2 \, \delta(t_a) \delta(t_b) \, M_{\sigma\tau}(a,b)}
 \\
 	& \quad - \sum_{\sigma=1}^n \int \de a \, \ththbar{a} \, \beta_g \, \delta(t_a) \, v\argp{r_\sigma(a)-w} \ .
\end{split}
\eeq
Using the replica-symmetric ansatz~\eqref{eq-memory-superkernel-explicit} for the memory superkernel (recall that we focus on the regime where the initial condition has a single
pure state), we can develop the SUSY formalism to write these expressions in terms of the original variables and obtain
\beq
\begin{split}
 S_{\text{eff}}^{(t)}
 =& \sum_{\sigma =1}^n \int_0^\infty \!\!\!\! \de t \, i\hat{r}_\sigma(t) \,
 		\arga{\int_0^\infty \!\!\!\! \de t' \, \argc{
 			\partial_{t'} \Gamma_R (t,t') + M_R(t,t')} r_\sigma(t')
 			- \nu_0(t) \, r_\sigma(t)
 			- v'\argp{r_\sigma(t)-w}
 			}
 \\
  & + \sum_{\sigma=1}^n \int_0^\infty \!\!\!\! \de t \int_0^\infty \!\!\!\! \de t' \, \frac12 \argc{\Gamma_C(t,t')+M_C^{(c)}(t,t')} \, i\hat{r}_\sigma(t) \, i\hat{r}_\sigma(t')
 \\
  &	+ \frac12 \int_0^\infty \!\!\!\! \de t \int_0^\infty \!\!\!\! \de t' \, \argp{\sum_{\sigma=1}^n i\hat{r}_\sigma(t)} M_d(t,t') \argp{\sum_{\tau=1}^n i\hat{r}_\tau(t')} \ ,
 \\
 S_{\text{eff}}^{(t,0)}
 =& \sum_{\sigma =1}^n \int_0^\infty \!\!\!\! \de t \, i\hat{r}_\sigma(t) \,
 			\argc{\Gamma_R^*(t,0) + \beta_g M_C^{(c)}(t,0)} r_\sigma(0)
	+ \int_0^\infty \!\!\!\! \de t \, \argp{\sum_{\sigma=1}^n i \hat{r}_\sigma(t)} \, \beta_g M_d(t,0) \, \argp{\sum_{\tau=1}^n r_\tau(0)} \ ,
 \\
 S_{\text{eff}}^{(0)}
 =& - \frac{\beta_g}{2} \argp{\lambda + \delta \nu(0) - \beta_g M_C^{(c)}(0,0)} \sum_{\sigma=1}^n r_\sigma(0)^2
  	- \sum_{\sigma=1}^n \beta_g \, v\argp{r_\sigma(0)-w}
	+ \frac{\beta_g^2}{2} M_d(0,0) \, \argp{\sum_{\sigma=1}^n r_\sigma(0) }^2 \ .
 \end{split}
\label{eq-time-dependent-effective-action}
\eeq

\subsubsection{Effective process}

We finally want to show that the time-dependent effective action ${ S_{\text{eff}}^{(t)} +  S_{\text{eff}}^{(t,0)}}$, 
written as in Eq.~\eqref{eq-time-dependent-effective-action}, can be derived from 
an effective Langevin dynamics in which each replica ${\sigma=1,\dots,n}$ evolves independently in presence of its own thermal noise,
$\z_\s(t)$, and of another independent noise $\zeta_d(t)$, common to all replicas, that represents the disorder. We now introduce this effective
process, and show that it leads to the same effective dynamical action as in Eq.~\eqref{eq-time-dependent-effective-action}.

We define the effective Langevin equation
\beq
\begin{split}
 \int_0^t  \de t' \, \Gamma_R(t,t') \, & \dot{r}_\sigma(t')
 = \int_0^t  \de t' \, M_R(t,t') \, r_\sigma(t')
  	- \underbrace{\big[ \hat{\nu}(t) + \delta \nu(t)\big]}_{=\tilde{\nu}(t)} r_\sigma(t) + \beta_g M_C^{(c)} \, (t,0)  r_\sigma(0)
  	- v' \argp{r_\sigma(t)-w}
  + \zeta_d(t) + \zeta_\s(t) \ ,
 \\
 \overline{\z_d (t)} &=0 \ , \quad
 \overline{\z_d (t) \z_d(t')}
 	= M_d(t,t') \ ,
 \\
 \moy{\zeta_\s (t)} &=0 \ , \quad
 \moy{\z_\s(t) \z_\t(t')} = \d_{\s\t} [ \G_C(t,t') + M_C^{(c)} (t,t') ] \ ,
\end{split}
\label{eq-SUSY-effective-process-dynamics}
\eeq
which is as expected the same effective noise and dynamics given in Sec.~\ref{sec-cavity-method-summary}, specifically in Eq.~\eqref{effective_spin2-bis}.
Note that we used the definition of $\nu_0(t)$ of Eq.~\eqref{eq-def-nu-0-scalar-SUSY}, and then performed an integration by part on $\partial_{t'}\Gamma_R(t,t')$ 
to get back the original friction term, and eliminate the term $\G_R^*(t,t)$ that appears in $\nu_0(t)$.
We can thus notice that the distinction between the singular and regular part of the friction kernel $\Gamma_R(t,t')$ is not relevant here, as expected.
It is straightforward to show, by computing the MSRJD action starting from this effective process, following the same derivation steps as from Eq.~\eqref{eq-Zdyn-very-first-version},
that one recovers ${ S_{\text{eff}}^{(t)} +  S_{\text{eff}}^{(t,0)}}$, as
written as in Eq.~\eqref{eq-time-dependent-effective-action}.

As for the initial condition encoded in ${S_{\text{eff}}^{(0)}}$, it also coincides with Eq.~\eqref{eq:cavity_prgivenz} obtained in Sec.~\ref{sec-cavity-method-summary}.
Once again, the most straightforward way to convince oneself that this is the case is to start from Eq.~\eqref{eq:cavity_prgivenz} obtained by the cavity.
Introducing replicas there in order to average over the `disorder' $\zeta_d(0)$, we obtain:
\beq
\begin{split}
 & \int_{\mathbb{R}} \de r(0) \int_{\mathbb{R}} \de \zeta_d(0) \, \mathcal{P} \argp{\zeta_d(0)} \, \mathcal{P} \argp{r(0) \vert \zeta_d(0)}
 \\
 &\stackrel{\eqref{eq:cavity_prgivenz}}{=}
 	\lim_{n \to 0} \int \argp{\prod_{\sigma=1}^n \de r_\sigma(0)} \,
 	\exp \arga{
	 	- \sum_{\sigma=1}^n \frac{r_\sigma(0)^2}{2(1-q_g)}
 		- \sum_{\sigma=1}^n \beta_g v\argp{r_\sigma (0) - w}
	}
  \times \overline{\exp\argp{\beta_g \zeta_d(0) \sum_{\sigma=1}^n r_\sigma(0)}}
 \\
 &= \lim_{n \to 0} \int \argp{\prod_{\sigma=1}^n \de r_\sigma(0)} \,
 	\exp \arga{
 	- \sum_{\sigma=1}^n \frac{r_\sigma(0)^2}{2(1-q_g)}
 	- \sum_{\sigma=1}^n \beta_g v\argp{r_\sigma (0) - w}
 	+ \frac12 \beta_g^2 \frac{q_g}{(1-q_g)^2} \argp{\sum_{\sigma=1}^n r_\sigma (0)}^2
	 }
 \\
 &=  \lim_{n \to 0} \int \argp{\prod_{\sigma=1}^n \de r_\sigma (0)} \, e^{S_{\text{eff}}^{(0)}} \ ,
\end{split}
\eeq
where we used for the last step the results in Sec.~\ref{sec-cavity-method-initial-condition} from the static cavity method, specifically Eqs.~\eqref{eq-tilde-lambda}, \eqref{eq-self-consistency-relations1} and~\eqref{eq-self-consistency-relations2}:
${1/(1-q_g) = \beta_g \tilde{\lambda}=\beta_g \argp{\lambda + \delta \nu (0) - \beta_g M_C^{(c)}(0,0)}}$
and ${q_g/(1-q_g)^2 = \beta_g^2 M_d(0,0)}$.

In summary, we have recovered via the saddle-point solution of the MSRJD action the same effective stochastic process as from the dynamical cavity method.
Physically what we did was to reformulate a problem of $N$ degrees of freedom ${\vec{X}(t)}$ in a quenched disorder ${\lbrace \vec{F}_\mu \rbrace}$ with the dynamical partition function
\beq
\begin{split}
 Z_{\mathrm{dyn}}
 &= \lim_{n\to 0} \overline{\moy{
 	\int \DD \vec X(t) \, {Z(\b_g)}^{n-1}e^{-\b_g \hat{\mathcal{H}} \argp{\vec X(0)}}
 	\prod_{i=1}^N \delta \argp{
 		\int_{0}^t  \!\! \de s \, \G_R(t,s) \, \dot{x}_i(s) - \hat \n(t) \, x_i(t) - \frac{\delta \mathcal{H} \argp{\vec{X}(t)}}{\delta x_i(t)} + \h_i(t)
 	}}}
\end{split}
\eeq
into a problem of a single-variable effective stochastic process capturing the mean-field properties in the thermodynamic limit:
\beq
\begin{split}
  \lim_{N \to \infty} Z_{\mathrm{dyn}}
 = \lim_{n \to 0} & \overline{ \left\langle
 	\int \DD \vec{r}(t) \, e^{\arga{
			-\frac12 \argp{\lambda + \delta \nu(0) - \beta_g M_C^{(c)}(0,0)} \sum_{\sigma=1}^n r_\sigma(0)^2
 			- \sum_{\sigma=1}^n \beta_g v(r_\sigma(0)-w)
 			- \beta_g \, \zeta_d(0) \, \sum_{\sigma=1}^n r_\sigma (0) 	
		}
 } 	\right.}
 \\
 & \quad \times
 	\prod_{\sigma=1}^n \delta \left(
		- \int_0^{\infty} \!\!\!\! \de t' \, \Gamma_R(t,t') \dot{r}_\sigma(t')
		- \argp{\hat{\nu}(t)+\delta\nu(t)} r_\sigma(t)
		- v'(r_\sigma (t) -w) + \zeta_d(t) + \zeta_\s(t)
 	\right.
 \\
 & \quad\quad\quad\quad\quad\quad\quad\quad
 	\left. \left. + \beta_g M_C^{(c)}(t,0) \, r_{\sigma}(0) + \int_0^{\infty} \!\!\!\! \de t' \, M_R(t,t') \, r_\sigma(t')
 	\phantom{\int_0^{\infty} \!\!\!\!\!\!\!\!} \right) \right\rangle
 \ .
\end{split}
\eeq
The interaction of the initial degrees of freedom with the initial quenched disorder has thus been replaced by the different memory kernel defined in Eq.~\eqref{eq-memory-superkernel-explicit}, as well as the correction ${\delta\nu(t)}$ defined in Eq.~\eqref{eq-correction-SUSY-Lagrange-explicit},  the Lagrange multipliers ${\hat{\nu}(t)}$ and ${\lambda = \hat{\nu}(0)}$.

\subsection{Dynamical equations for the correlation and response functions}
\label{sec-SUSY-dynamics}

Having recovered the same effective stochastic process as in Sec.~\ref{sec-cavity-method-summary}, we finally need to derive the dynamical equations for the correlation and response functions, as well as for the Lagrange multiplier.
These can be found starting from the closure relation Eq.~\eqref{eq:HSSUSY3}.

Similarly to the memory superkernel in Eq.~\eqref{eq-memory-superkernel-explicit},
we can use the following replica-symmetric (RS) ansatz for the saddle-point equation:
\beq
\begin{split}
\widetilde{Q}_{\s\t}(a,b)
 &= \underbrace{\argc{C(t_a,t_b) + \ththbar{a} R(t_b,t_a) + \ththbar{b} R(t_a,t_b) }}_{\equiv \hat Q(a,b)} \,  \delta_{\s\t}
	+ (1-\d_{\s\t})C_d(t_a,t_b)
\end{split}
\eeq
and we will treat separately the scalar, $\ththbar{a}$ and $\ththbar{b}$ contributions of the following SUSY equation that derives from 
Eq.~\eqref{eq:HSSUSY3}:
\beq
\begin{split}
 \argp{\ththbar{a} + \ththbar{b}} \delta_{\sigma \tau} \, \delta(t_a-t_b)
 =	& - \sum_{\gamma=1}^n \int \de c \, \arga{
		\hat{\Gamma}(a,c) \, \delta_{\sigma \gamma}
		+ \mathcal{M}(a,c) \, \delta_{\sigma \gamma}
		+ M_d(t_a,t_c)
	 	} \, \widetilde{Q}_{\gamma \tau}(c,b)
 \\ & + \argc{\hat{\nu}(t_a) + \delta \nu(t_a) + \Gamma_R^*(t_a,t_a)} \, \widetilde{Q}_{\sigma \tau}(a,b)
 \\ & - \sum_{\gamma =1}^n \int \de c \,
		\argc{\ththbar{a} \, \delta (t_a) \, \Gamma_R^*(t_c,t_a) + \ththbar{c} \, \delta (t_c) \, \Gamma_R^*(t_a,t_c)} \delta_{\sigma \gamma} \, \widetilde{Q}_{\gamma \tau}(c,b)
 \\ & - \sum_{\gamma =1}^n \int \de c \,
		\argc{\ththbar{a} \, \delta (t_a) + \ththbar{c} \, \delta (t_c)} \beta_g \argc{\mathcal{M}(a,c) \, \delta_{\sigma \gamma} + M_d(t_a,t_c)} \widetilde{Q}_{\gamma \tau}(c,b)
 \\ & - \sum_{\gamma =1}^n \int \de c \,
		\ththbar{a} \ththbar{c} \, \delta(t_a) \delta(t_c) \, \beta_g^2 \argc{M_C^{(c)}(0,0) \, \delta_{\sigma \gamma} + M_d(0,0)} \, \widetilde{Q}_{\gamma \tau}(c,b)
 \\ & + \ththbar{a} \, \delta (t_a) \, \beta_g(\lambda + \delta \nu (0)) \, \widetilde{Q}_{\sigma \tau}(a,b) \ .
\end{split}
\eeq
Starting from the scalar case, we obtain a first equation for ${\sigma = \tau}$:
\beq
\begin{split}
 \int_0^t \de s \, \Gamma_R (t,s) \, \partial_s C(t',s)
 =	&	- \argc{\hat{\nu}(t) + \delta \nu(t)} \, C(t,t')
 		+ \int_0^{t'} \de s \, \argc{\Gamma_C(t,s) + M_C^{(c)}(t,s) + M_d(t,s)} \, R(t',s)
 \\
 	&	+ \int_0^t \de s \, M_R(t,s) \, C(t',s)
 		+ \beta_g M_C^{(c)}(t,0) \, C(0,t')
 \\
 	&	+ \beta_g M_d(t,0) \, \argc{C(0,t') + (n-1) \, C_d(0,t')} \ ,
\end{split}
\label{eq-SUSY-dyn-correlation}
\eeq
and a second equation for ${\sigma \neq \tau}$:
\beq
\begin{split}
 \int_0^t \de s \, \Gamma_R (t,s) \, \partial_s C_d(t',s)
 =	&	- \argc{\hat{\nu}(t) + \delta \nu(t)} \, C_d(t,t')
 		+ \int_0^{t'} \de s \, M_d(t,s) \, R(t',s)
 \\
 	&	+ \int_0^t \de s \, M_R(t,s) \, C_d(t',s)
 		+ \beta_g M_C^{(c)}(t,0) \, C_d(0,t')
 \\
 	&	+ \beta_g M_d(t,0) \, \argc{C(0,t') + (n-1) \, C_d(0,t')} \ .
\end{split}
\label{eq-SUSY-dyn-correlation-bis}
\eeq
Once we take the limit ${n \to 0}$, that is required for the computation of $Z_{\rm dyn}$ according to Eq.~\eqref{eq-Zdyn-very-first-version}, we recover as expected the same dynamical equations for the correlation functions as via the cavity approach, see Eqs.~\eqref{Ccavity}-\eqref{Ccavity-bis}.
Consequently the Lagrange multiplier is given by the same equation as Eq.~\eqref{Lag_eq}, which was itself deduced from Eq.~\eqref{Ccavity} by imposing that ${C(t,t)=1}$. 

Next, we consider the $\ththbar{b}$ contribution, recalling that by causality $\G_R^*(t,s) \propto \theta(t-s)$:
\beq
\begin{split}
 &	\argc{- \int_{0}^{t} \!\!\!\! \de s \, \partial_s \Gamma_R(t,s) \, R(s,t')
 	+ \Gamma_R^*(t,t) \, R(t,t') - \Gamma_R^*(t,0) \, R(0,t')} \delta_{\sigma \tau}
 \\	
 & = \arga{\delta(t-t')
 	+ \int_{0}^{\infty} \!\!\!\! \de s \, M_R(t,s) \, R(s,t')
 	- \argc{\hat{\nu}(t) + \delta \nu(t)} \, R(t,t')} \delta_{\sigma \tau}
 	+ \beta_g \argc{M_C^{(c)}(t,0) \, \delta_{\sigma \tau} + M_d(t,0)} \, R(0,t') \ .
\end{split}
\eeq
Note that this equation makes sense only for $t>t'>0$, and one can then send $t'\to 0^+$. Hence
for different replica indices ${\sigma \neq \tau}$ we simply have that
${\beta_g M_d(t,0) \, R(0,t') = 0}$, because ${R(s,t') \propto \theta(s-t')}$ by causality.
Consequently, also for ${\sigma = \tau}$ all the terms involving $R(0,t')$ disappear.
Integrating by parts the first line, recalling that in the original convention the integrals are intended up to $t^+$,
we recover as expected the same dynamical equation for the response function as via the cavity method in Eq.~\eqref{Rcavity}, for ${t \geq t'}$:
\beq
\begin{split}
 \int_{t'}^t \de s \, \Gamma_R (t,s) \, \partial_s R(s,t')
 =	\delta(t-t')
 	&	- \argc{\hat{\nu}(t) + \delta \nu(t)} \, R(t,t')
 		+ \int_{t'}^{t} \de s \, M_R(t,s) \, R(s,t') \ .
\end{split}
\label{eq-SUSY-dyn-response}
\eeq
Note that an equation for $R(t,t')$ can be equivalently derived from the $\ththbar{a}$ contribution; showing that the two equations coincide
requires however some manipulations.

In order to make a connection with the dynamical equations given in Ref.~\cite{BK13}, we introduce the following notations:
\beq
\begin{cases}
 D(t,s) & \equiv \Gamma_C^*(t,s) + M_C^{(c)}(t,s) + M_d(t,s) =  \Gamma_C^*(t,s) + M_C(t,s) \ ,
 \\
 \Sigma (t,s) & \equiv \partial_s \Gamma_R^*(t-s) + M_R(t,s)
 					= - \partial_t \Gamma_R^*(t-s) + M_R(t,s) \ ,
 \\
 \mu (t) & \equiv \hat{\nu}(t) + \delta \nu (t) + \Gamma_R^*(t,t)
			 = \hat{\nu}(t) + \delta \nu (t) + \Gamma_R^*(0) \ . 
\end{cases}
\eeq
Using these definitions, we can rewrite the set of dynamical equations as
\begin{eqnarray}
&& \begin{split}
 \gamma_R \partial_{t} C(t',t)
 = & \int_0^{t'} \de s \, D(t,s) R(t',s) + 2T \gamma_C \, R(t',t)
 	+ \int_0^{t} \de s \, \Sigma (t,s) C(t',s)
 	- \mu(t) \, C(t,t')
 \\ &  \quad + \argc{\beta_g M_C(t,0) + \Gamma_R^*(t,0)} C(t',0)
 		- \beta_g M_d(t,0) C_d(0,t')
 \ ,
 \end{split}
\\
&&  \begin{split}
 \gamma_R \partial_{t} C_d(t',t)
 = & \int_0^{t'} \de s \, M_d(t,s) R(t',s)
 	+ \int_0^{t} \de s \, \Sigma (t,s) C_d(t',s)
 	- \mu(t) \, C_d(t,t')
 \\ &  \quad + \argc{\beta_g M_C^{(c)}(t,0) + \Gamma_R^*(t,0)} C_d(t',0)
 		+ \beta_g M_d(t,0) \argc{C(0,t')-C_d(0,t')}
 \ ,
 \end{split}
\\
&& \gamma_R \partial_t R(t',t)
 = \delta(t-t') + \int_{t'}^t  \de s \, \Sigma (t,s) R(s,t') - \mu(t) \, R(t,t')
 \ ,
\\
&& \begin{split}
 \mu(t)
 = & \int_0^{t} \de s \, \argc{D(t,s) R(t,s) + \Sigma(t,s) C(s,t)}
 	- \gamma_R \partial_t C(t,t^+)
 \\ & \quad\quad\quad\quad + \argc{\beta_g M_C(t,0) + \Gamma_R^* (t,0)} C(t,0)
 	- \beta_g M_d(t,0) \, C_d(t,0)
  \ .
 \end{split}
\end{eqnarray}
Compared to the dynamical equations for the driven $p$-spin presented in Ref.~\cite{BK13}, these equations present
additional terms due to the memory of the initial condition, 
along with the expressions of the memory kernels and ${\delta \nu(t)}$ which differ of course 
between the driven perceptron and the driven $p$-spin.

This concludes our alternative derivation of the effective stochastic process and dynamical equations for the thermodynamic limit of the perceptron model, obtained here via the saddle point of the SUSY path integral and previously via the cavity method.
We emphasize that a summary of the dynamical equations for the effective stochastic process has been given
in section~\ref{sec-cavity-method-summary}.

\section{Conclusions}
\label{sec-conclusion}

In this paper, we have derived the dynamical mean field equations (DMFE) that describe the Langevin dynamics for the random continuous perceptron model.
We have used two distinct approaches.
First we have developed a dynamical cavity approach to construct the single-variable effective process from which we extracted the equations for the correlations and response functions.
Second we have developed a path integral approach in its supersymmetric formalism and we have shown that it leads to the same results as the cavity method.
The final effective process is a stochastic equation with memory and correlated Gaussian noise that must be determined self-consistently.

In our derivations we have assumed \emph{generic} friction and noise kernels ${\Gamma_C}$ and ${\Gamma_R}$, and a stochastic initial condition which can be tuned by $\beta_g$ (for instance ${\beta_g=0}$ corresponds to a uniform initial condition), so that our DMFE can broadly describe out-of-equilibrium settings in the thermodynamic limit.
Consequently, this work paves the way to several further developments, a few of them being listed thereafter:
\begin{itemize}

\item
The dynamical cavity method could be used to give a more straightforward derivation of the dynamics of sphere systems in high dimensions \cite{MKZ16a, MKZ16b}.
A first attempt in this direction has been made in Ref.~\cite{Sz17}.
However in that work, the equivalent of Eq.~\eqref{effective_spin2} was not fully derived microscopically.
This is an important missing step that remains to be made, and our results could provide useful inspiration.

\item
The solution we have developed is very general.
In particular it includes the possibility to have an active drive in the dynamics.
This could be useful to study with a purely microscopic approach simple models of active matter and in particular how the active drive interplays with structural or quenched disorder \cite{BK13}.

\item
In neural neworks the construction of the solution of optimal synaptic weights is made using a series of algorithmic procedures that aim at minimising the loss function or Hamiltonian.
In this paper we have analyzed the simplest algorithm to find local optimisers that is a simple gradient descent (in presence or not of stochastic noise).
It could be interesting to investigate how this approach, especially in its cavity version, can be developed to study more complicated minimization algorithms.

\item
The perceptron model is a general model for glasses.
In addition it includes a jamming transition.
The critical dynamics close to jamming has never been studied analytically, but many numerical results are available~\cite{OT07,Ol10,DDLW15,KCIB15}.
The methods developed here could be used to analytically reproduce at least part of these results.
Moreover, the aging dynamics in the vicinity of jamming could be somewhat different from the standard solution of aging in disordered spin glass models~\cite{CK93}, because the vicinity to an isostatic point could strongly affect the response of the system to abrupt changes in the external parameters \cite{FS17}.

\end{itemize}

Finally we note that the cavity approach we have developed can be used to study analytically the molecular dynamics version of the model \cite{CLN17}.
Indeed it is quite straightforward to switch off the thermal noise and introduce an inertial term in the dynamical equations (of the type of $m \ddot x_i$).
The equations for the final effective stochastic process can be obtained using the same steps we described in this work.
This corresponds to study molecular dynamics with a stochasticity included in the initial conditions.

It should be noted that the DMFE equations we derived are formulated in terms of an effective one-dimensional stochastic process with colored noise.
The noise kernel, however, is not known analytically.
It is instead expressed self-consistently as a correlation function of the stochastic process itself.
On the one hand, this makes it very challenging to obtain analytic solutions to the DMFEs, which therefore have \textit{a priori} to be solved numerically.
On the other hand, this self-consistent determination of the kernel highlights a fundamental difference between glassy systems and other disordered systems such as in the depinning of elastic lines moving in a quenched disordered environment~\cite{LeDW13,Papa16}.
While in depinning the noise is determined by the quenched environment, leading to effective mean field equations with fixed noise kernels, as e.g. in the case of ABBM models~\cite{LeDW13}, in glassy systems the noise is self-consistently determined by the interactions, leading to the self-consistency condition for the noise kernel that we found in the DMFEs.
It would be certainly very interesting to explore the analogies and differences between these two kinds of systems in more details, in particular to explore whether the DMFEs reduce, in some well-defined limit, to simpler equations such as the ABBM model.


\section*{Acknowledgments}

We would like to thank Jorge Kurchan, Thibaud Maimbourg and Grzegorz Szamel for fruitful discussions related to this work.
This work was supported by a grant from the Simons Foundation ($\sharp$454935, Giulio Biroli; $\sharp$454955, Francesco Zamponi),
and by ``Investissements d'Avenir'' LabEx PALM (ANR-10-LABX-0039-PALM) (Pierfrancesco Urbani).




\end{document}